\documentclass[ste,twoside]{stefano}

\usepackage{amsmath}
\usepackage{amssymb}
\usepackage{graphics}
\usepackage{rotating}
\usepackage{cite}
\usepackage{color}

\def\makeheadbox{{%
\hbox to0pt{\vbox{\baselineskip=10dd\hrule\hbox
to\hsize{\vrule\kern3pt\vbox{\kern3pt
\hbox{  {\sc Quaternionic potentials in NRQM} }
\hbox{ Journal of Physics A {\bf 35}, 5411-5426 (2002)
\hspace*{7.3cm} 
{\color{blue}{$\boldsymbol{\Sigma \delta \Lambda}$}} }
\kern3pt}\hfil\kern3pt\vrule}\hrule}%
\hss}}}

\def\r{\boldsymbol{r}}
\def\Z{\mbox{\tiny $Z$}}
\def\V{\mbox{\tiny $V$}}
\def\W{\mbox{\tiny $W$}}
\def\Wm{\mbox{\tiny $|W|$}}
\def\te{\mbox{\tiny $\theta$}}

\def\+{\mbox{\tiny $\dag$}}
\def\0{\mbox{\tiny $0$}}
\def\1{\mbox{\tiny $1$}}
\def\2{\mbox{\tiny $2$}}
\def\3{\mbox{\tiny $3$}}
\def\4{\mbox{\tiny $4$}}
\def\5{\mbox{\tiny $5$}}
\def\6{\mbox{\tiny $6$}}
\def\7{\mbox{\tiny $7$}}
\def\8{\mbox{\tiny $8$}}
\def\9{\mbox{\tiny $9$}}
\def\a{\mbox{\tiny $a$}}
\def\b{\mbox{\tiny $b$}}

\def\t{\mbox{\tiny $t$}}

\def\e{\mbox{\tiny $\eta$}}

\def\th{\mbox{\tiny $\theta$}}

\def\x{\mbox{\tiny $x$}}

\def\tot{\mbox{\tiny $\to$}}

\def\pem{\mbox{\tiny $\pm$}}
\def\mi{\mbox{\tiny $-$}}
\def\pl{\mbox{\tiny $+$}}
\def\={\mbox{\tiny $=$}}
\def\T{\mbox{\tiny T}}
\def\L{\mbox{\tiny $L$}}
\def\R{\mbox{\tiny $R$}}

\def\Co{\mbox{\tiny $\mathbb{C}$}}

\def\E{\mbox{\tiny $E$}}
\begin{document}
%

\title{QUATERNIONIC POTENTIALS IN\\NON-RELATIVISTIC QUANTUM MECHANICS}

\author{
Stefano De Leo\inst{1}
\and  
Gisele C. Ducati\inst{1,2}
\and
Celso C. Nishi\inst{3}
}

\institute{
Department of Applied Mathematics, State University of Campinas\\ 
PO Box 6065, SP 13083-970, Campinas, Brazil\\
{\em deleo@ime.unicamp.br}\\
{\em ducati@ime.unicamp.br}
\and
Department of Mathematics, University of Parana\\
PO Box 19081, PR 81531-970, Curitiba, Brazil\\
{\em ducati@mat.ufpr.br}
\and
Department of Cosmic Rays and Chronology, State University of Campinas\\
PO Box 6165, SP 13083-970, Campinas, Brazil\\
{\em ccnishi@ifi.unicamp.br}
}


\date{{\em June 21, 2002}}

\abstract{We discuss the Schr\"odinger equation in presence of quaternionic 
potentials. The study is performed analytically as long as it 
proves possible, when not, we resort to numerical calculations.
The results obtained could be useful to investigate an underlying 
quaternionic quantum dynamics in particle physics. Experimental tests and
proposals to observe quaternionic quantum effects by neutron interferometry
are briefly reviewed.
}








\titlerunning{Quaternionic potentials in NRQM}
 
\maketitle


\section*{I. INTRODUCTION}

After the classical mathematical and physical works on  
foundations of quaternionic quantum mechanics~\cite{KAN60,FIN62,FIN63,EMC63}, 
there has been, in recent years, a widely interest in formulating quantum 
theories by using the non commutative ring of 
quaternions~\cite{HOR84,ADL85,ADL86,ADL86b,ADL88,ADL96,ROT89,DEL92}. 
Some of the main results coming out from the use of {\em new} algebraic
structures in particle physics are reviewed in the books of 
Dixon~\cite{DIX} and G\"ursey~\cite{GUR}. For a detailed discussion of 
quaternionic quantum mechanics and field theory we quote the excellent book 
of Adler~\cite{ADL}.

The present paper has grown from an attempt to understand the 
experimental proposals~\cite{PER79,KAI84,KLE88} and theoretical 
discussions~\cite{DAV89,DAV90,DAV92} underlying the quaternionic 
formulation of the Schr\"odinger equation. The main difficulty in obtaining 
quaternionic solutions of physical problem is due to the fact that, in 
general, the standard mathematical methods of resolution break down.
In the last years, some of these problems have been overcome. In particular,
the discussion of quaternionic eigenvalue equations~\cite{RCEE} and 
differential operators~\cite{QDO} is {\em now} recognized quite satisfactory.
On the other hand, physical interpretations of quaternionic solutions
represent a more delicate question~\cite{ADL}. In discussing the
Schr\"odinger equation what is still lacking is to understand the
role that quaternionic potentials could play in quantum mechanics and
where deviations from the standard theory would appear.

The earliest experimental proposals to test quaternionic deviations
from complex quantum mechanics were made by Peres~\cite{PER79} 
who suggested that
 the  non commutativity of  quaternionic phases could be observed in 
Bragg scattering by crystal made of three different 
atoms, in neutron interferometry and in meson regeneration.
In 1984, the neutron interferometric experiment was realized by Kaiser,
George and Werner~\cite{KAI84}. 
The neutron wave function traversing  slabs of two dissimilar materials 
(titanium and aluminum) 
should experience  the non commutativity of the phase shifts
when the order in which the barriers are traversed is reversed. 
The experimental result showed that the phase shifts 
commute to better than one part in $3 \times 10^{\4}$.
To explain this null result, Klein postulated~\cite{KLE88}  
that quaternionic
potentials act only for some of the fundamental forces and 
proposed an experiment for testing possible violations
of the Schr\"odinger equation by permuting the order in 
which nuclear, magnetic and gravitational potentials act on neutrons 
in an interferometer.  

The first theoretical analysis of two 
quaternionic potential barriers 
was developed by Davies and McKellar~\cite{DAV92}. In their paper, 
by translating the   quaternionic Schr\"odinger equation into a pair of 
coupled complex equations
and solving the corresponding complex system by numerical methods, 
Davies and McKellar showed that,   
notwithstanding the presence of complex {\em instead} of quaternionic phases,
the predictions of quaternionic quantum mechanics  
differ from those of the usual theory. In particular, they pointed out that 
differently from the  complex quantum 
mechanics prediction, where the left and right transmission amplitudes, 
$t_{\L}$ and $t_{\R}$, are equal in magnitude and in phase,  in the 
quaternionic quantum mechanics only the magnitudes $|t_{\L}|$ and 
$|t_{\R}|$ are equal. So, the measurement 
of a phase shift should be an indicator of quaternionic effects {\em and}
of space dependent phase potentials. However, this conclusion leads to the
embarrassing question of why there was no phase change in the experiment 
proposed by Peres and realized by Kaiser, George and Werner. To reconcile the
theoretical predictions with the experimental observations, Davies and McKellar
reiterated the Klein conclusion and suggested to subject the neutron beam to 
different interactions in permuted order. In the final chapter of the Adler 
book~\cite{ADL}, we find and intriguing question. Do the Kayser and 
colleagues experiment, and the elaborations on it proposed by Klein actually
test for residual quaternionic effects? According to the non relativistic 
quaternionic scattering theory developed by Adler~\cite{ADL} the answer is 
clearly no. Experiments to detect a phase shift are equivalent to detect 
time reversal violation, which so far has not been detectable in 
neutron-optical experiments.

In this paper, after a brief introductory discussion of quaternionic  
anti-self-adjoint operators, stationary states and time reversal invariance, 
we study the phenomenology of quaternionic one-dimensional square potentials.
The $j$-$k$ part of these potentials is treated as a perturbation of the 
complex case. We show that there are many possibilities in looking for
quaternionic deviations from the standard (complex) theory. Nevertheless, in
particular cases, we have to contend with quaternionic effects which 
minimize the deviations from complex quantum mechanics. With this 
paper, we would like to close the debate on the role that quaternionic 
potentials could play in quantum mechanics, but more realistically, we simply
contribute to the general discussion.


\section*{II. QUATERNIONIC SCHR\"ODINGER EQUATION}

In the standard formulation of non-relativistic 
quantum mechanics, the 
complex wave function   $\varphi(\r ,t)$, describing a particle without spin 
subjected to the influence of a 
real potential $V(\r ,t)$, satisfies the Schr\"odinger equation 
\begin{equation}
\label{se}
\partial_{\t} \varphi( \r,t) = 
\mbox{$\frac{i}{\hbar}$} \, \left[ \, 
\mbox{$\frac{\hbar^{\2}}{2m}$} \, \nabla^{\2} 
- V(\r,t) \, \right] \, \varphi(\r,t)~.
\end{equation}
In quaternionic quantum mechanics~\cite{ADL}, the anti-self-adjoint  
operator
\[
\mathcal{A}^{\V}(\r , t ) = \mbox{$\frac{i}{\hbar}$} \, \left[ \, 
\mbox{$\frac{\hbar^{\2}}{2m}$} \, \nabla^{\2} 
- V(\r,t) \, \right]
\]
can be generalized by introducing the complex 
potential  $W(\r ,t) = 
\left| W(\r ,t) \right| \, \exp [ i \theta (\r,t)]$, 
\[
\mathcal{A}^{\V , \W}(\r , t) = \mbox{$\frac{i}{\hbar}$} \, \left[ \, 
\mbox{$\frac{\hbar^{\2}}{2m}$} \, \nabla^{\2} 
- V(\r,t) \, \right]
+ \mbox{$\frac{j}{\hbar}$} \, W(\r ,t)~.
\]
The anti-hermiticity is required to 
guarantee the time conservation of  transition probabilities. As a 
consequence of this generalization for the  anti-self-adjoint Hamiltonian   
operator, the quaternionic wave function  $\Phi(\r ,t)$ 
satisfies the following equation 
\begin{equation}
\label{qse}
\partial_{\t} \Phi( \r,t) = \left\{ \, 
\mbox{$\frac{i}{\hbar}$} \, \left[ \, 
\mbox{$\frac{\hbar^{\2}}{2m}$} \, \nabla^{\2} 
- V(\r,t) \, \right] 
+ \mbox{$\frac{j}{\hbar}$} \, W(\r ,t) \, \right\}
\, \Phi(\r,t)~.
\end{equation}
Exactly as in the case of the standard quantum mechanics, 
we can define a current density
\[
\boldsymbol{J} =   
\mbox{$\frac{\hbar}{2m}$} \, \left[ \, 
\left( \boldsymbol{\nabla} \overline{\Phi} \right)  
\, i \, \Phi -  
\overline{\Phi}  \, i \, \boldsymbol{\nabla}  \Phi \, \right]
\]
and a probability density
\[ \rho = \overline{\Phi} \Phi~.\]  
Due to the non commutativity nature of quaternions, the position of the 
imaginary unit $i$ 
in the current density is fundamental
to obtain the continuity equation 
\begin{equation}
\label{con}
\partial_{\t} \rho + 
\boldsymbol{\nabla} \cdot \boldsymbol{J} = 0~.
\end{equation}

\subsection*{A. STATIONARY STATES}

The quaternionic
Schr\"odinger equation in presence of   time-independent
potentials
\[ \left[ \, V(\r) \, , \, | W (\r) | \, , \, \theta (\r) \, 
\right] 
\]
reads
\begin{equation}
\label{qset}
\partial_{\t} \Phi( \r,t) = \left\{ \, 
\mbox{$\frac{i}{\hbar}$} \, \left[ \, 
\mbox{$\frac{\hbar^{\2}}{2m}$} \, \nabla^{\2} 
- V(\r) \, \right] 
+ \mbox{$\frac{j}{\hbar}$} \, W(\r) \, \right\}
\, \Phi(\r,t)~.
\end{equation}
The quaternionic stationary state wave function
\[
\Phi(\r, t) = \Psi(\r) \, 
\exp[ \, - \, \mbox{$\frac{i}{\hbar}$} \, E \, t \, ]
\]
is solution of Eq.~(\ref{qset}) on the condition that
$\Psi(\r)$ be solution of the time-independent Schr\"o\-din\-ger equation 
\begin{equation}
\label{qse2} 
\left[ \,  
i \, \mbox{$\frac{\hbar^{\2}}{2m}$} \, \nabla^{\2} 
- i \, V(\r)  + j \, W(\r) \, \right] \, \Psi(\r) +  
\Psi(\r) \, i \, E = 0~.
\end{equation}
Eq.~(\ref{qse2}) represents a right complex eigenvalue equation
on the quaternionic field~\cite{RCEE}
\[
\mathcal{A}^{\V , \W}_{\E}(\r ) \, \Psi(\r) = - \, \Psi(\r) \, i E~.
\]
The allowed
energies are determined by the right complex eigenvalues 
$\lambda=i \, E$ 
of the  quaternionic  linear anti-self-adjoint operator 
$\mathcal{A}_{\E}^{\V , \W}(\r)$.
The stationary state wave functions are particular solutions of
Eq.~(\ref{qset}). More general solutions can be constructed by 
superposition of such particular solutions. Summing over various allowed
values of $E$, we get 
\begin{equation}
\label{sol}
\Phi ( \r , t ) = \sum_{\E} \, \Psi(\r) \, 
\exp[ \, - \, \mbox{$\frac{i}{\hbar}$} \, E \, t \, ] \, 
q_{\E}~,
\end{equation}
where $q_{\E}$ are constant quaternionic coefficients. 
The summation may imply an 
integration if the energy spectrum of $E$ is continuous.

\subsection*{B. TIME REVERSAL INVARIANCE}

From Eq.~(\ref{qset}), we can immediately obtain  the time-reversed 
Schr\"odinger equation
\begin{equation}
\label{tr}
\partial_{\t} \Phi_{\T}( \r, - t) = - \left\{ \, 
\mbox{$\frac{i}{\hbar}$} \, \left[ \, 
\mbox{$\frac{\hbar^{\2}}{2m}$} \, \nabla^{\2} 
- V(\r) \, \right] 
+ \mbox{$\frac{j}{\hbar}$} \, W(\r) \, \right\}
\, \Phi_{\T}(\r, - t)~.
\end{equation}
In complex quantum mechanics the $*$-conjugation yields a time-reversed 
version  of the original Schr\"odinger equation.  
In quaternionic quantum mechanics there does {\em not} exist
a universal time reversal operator~\cite{ADL}. 
Only a {\em restricted} class of time-independent 
quaternionic potentials, i.e. 
\[
W(\r)=|W(\r)| \, \exp [i \theta]~,
\]
is time reversal invariant.  For these potential, 
\begin{equation}
\Phi_{\T}(\r, - t) =  u \,   \Phi(\r, t) \, \bar{u}~,~~~
u = k \, \exp \left[ i \, \theta \right]~.
\end{equation}
For complex wave functions, we recover the standard result
$\Phi_{\T}(\r, - t) =  \Phi^{*}(\r, t)$.

\subsection*{C. ONE-DIMENSIONAL SQUARE POTENTIALS}

In solving the quaternionic Schr\"odinger equation,  
a great mathematical simplification results 
from the assumption that the wave function and the potential 
energy depend only on the $x$-coordinate,
\begin{equation}
\label{qse3} 
i \, \mbox{$\frac{\hbar^{\2}}{2m}$} \, \ddot{\Psi}(x) +  
\left[ \, j \, W(x) - i \, V(x) \, \right] \, \Psi(x) +  
\Psi(x) \, i \, E = 0~.
\end{equation}
We shall consider one-dimensional problems with a potential which is  pieced 
together from a number of {\em constant} portions, i.e. square potentials. 
In the
potential region 
\[ [ \, V \, ; \, |W| \, , \, 
\theta \, ]
\] 
the solution of the second order differential 
equation~(\ref{qse3}) is given by~\cite{QDO}
\begin{eqnarray}
\Psi(x) & =  & u^{\E;\Wm,\te} \,  \left\{ \exp \left[ 
\, z_{\mi}^{\E ; \V , \Wm}  \, x \, \right] \, 
c_{\1} + 
\exp \left[ \, - \, z_{\mi}^{\E ; \V , \Wm}  \, x \, \right] \, 
c_{\2} \right\} + \nonumber \\
 & & 
v^{\E ; \Wm,\te} \,  \left\{ \exp \left[ \, z_{\pl}^{\E ; \V , \Wm} \, 
x \, \right] \, 
c_{\3} + 
\exp \left[ \, - \, z_{\pl}^{\E ; \V , \Wm} \, x \, \right] \, 
c_{\4} \right\}~,
\end{eqnarray}
where $c_{\1,...,\4}$ are complex coefficients determined by the 
boundary conditions,
\[ 
z_{\pem}^{\E ; \V , \Wm} = 
 \mbox{\footnotesize{$ \sqrt{ \frac{2m}{\hbar^{\2}} \, \left( 
V \pm \sqrt{E^{\2} - |W|^{\2}} \right)}$}} ~\in \mathbb{C}(1,i)
\]
and
\[
u^{\E;\Wm,\te} = 
 \mbox{\footnotesize{$ 
\left( 1 - k \, \frac{|W| \, \exp[i \theta]}{E + \sqrt{E^{\2} - |W|^{\2}}^{} }
\right)$}}~,~~~
v^{\E; \Wm,\te} = 
 \mbox{\footnotesize{$
\left( j - \frac{i \, |W| \, \exp[- 
i \theta]}{E + \sqrt{E^{\2} - |W|^{\2}}^{}}\right)
$}}~\in \mathbb{H}~.
\]
In the free potential region, the solution 
reduces to
\begin{eqnarray*}
\Psi(x) & = & \exp \left[ \, i \,  
\mbox{$\frac{p}{\hbar}$} \, x \, \right] \, c_{\1} +
\exp \left[ \, -  i \,  
\mbox{$\frac{p}{\hbar}$} \, x \, \right] \, c_{\2} + j \, \left\{ 
\exp \left[ \,  
\mbox{$\frac{p}{\hbar}$} \, x \, \right]
\, c_{\3} +
 \exp \left[ \, -   
\mbox{$\frac{p}{\hbar}$} \, x \,  \right] \, c_{\4}
\right\}~,
\end{eqnarray*}
where $p = \sqrt{2mE}$. 
For scattering problems with a wave function incident from the left 
on quaternionic potentials, we have 
\begin{equation}
\label{psii}
\Psi_{\mi}(x) = 
\exp [ \,  i \,  
\mbox{$\frac{p}{\hbar}$} \, x \, ] + r \, \exp[ \, - i \,  
\mbox{$\frac{p}{\hbar}$} \, x \, ] + 
j \, \tilde{r} \, \exp[ \,  
\mbox{$\frac{p}{\hbar}$} \, x \,]~,
\end{equation}
where $|r|^{\2}$ is the standard probability of reflection and 
$|\tilde{r} \, \exp[  \,  
\mbox{$\frac{p}{\hbar}$} \, x \, ]|^{\2}$ represents an additional  evanescent 
reflection.


\section*{III. TIME REVERSAL INVARIANT (TRI) POTENTIAL BARRIER}

Let us consider the TRI potential 
\[ 
[ \, V(x) \, ; \, |W(x)| \, , \, \theta\, ]~.
\] In Eq.~(\ref{qse3}),  
the space-independent phase $\theta$ can be removed by taking the    
transformation 
\begin{equation}
\label{tran}
\Psi(x) \rightarrow  
\exp \left[i \,  \mbox{$\frac{\theta}{2}$} \right] \, \Psi(x) \,  
\exp \left[ - i \,  \mbox{$\frac{\theta}{2}$} \right]~.
\end{equation} 
Under this transformation
\[ 
u^{\E;\Wm,\te} \to u^{\E;\Wm}~~~\mbox{and}~~~
v^{\E;\Wm,\te} \to v^{\E;\Wm} \, \exp[-i \, \theta]~.
\]
Reflection and transmission probabilities do {\em not} change
(actually the exponential  $\exp[-i \, \theta]$ can be absorbed in 
the complex coefficients $c_{\3 , \4}$). So, without lost 
of generality, we can discuss the quaternionic  Schr\"odinger equation
in presence of the square potential 
\[ \left[ \, V(x) \, ; \, |W(x)| \, \right] \]
which has the following shape
\begin{center}
{\color{blue}{
\begin{tabular}{ccccccccccccc}
& &  & & & & &  & \\
~{\color{black}{{\sc Region I}$_{\mi}$}} &   & 
{\sc Region II}$_{\mi}$ &  & 
{\color{red}{{\sc Region III}}} &   & 
{\sc Region II}$_{\pl}$ &   & 
{\color{black}{{\sc Region I}$_{\pl}$}} \\ 
& &  & & & & &  & \\
{\color{black}{$\left[ \, 0 \, ; \, 0  \, \right]$}}     &  & 
$\left[ \, V \, ; \, 0 \,  \right]$            &  & 
$\left[ \, V \, ; \, {\color{red}{W}} \,  \right]$            &  & 
$\left[ \, V \, ; \, 0 \,  \right]$            &  & 
{\color{black}{$\left[ \, 0 \, ; \, 0 \,  \right]$}} \\
& &  & & & & &  & \\
\hline
& $^{|}$ & 
& $^{|}$ & 
& $^{|}$  & 
&  $^{|}$ & \\
& $-a$ & 
& $-b$ & 
& $+b$  & 
&  $+a$ & \\
\end{tabular}
}}
\end{center}
The particle is free for $x<-a$, where the solution is given by 
Eq.~(\ref{psii}), 
and for $x>a$, where the solution is
\begin{equation}
\label{psit}
\Psi{\pl}(x) = 
t \, \exp[ \, i \,  
\mbox{$\frac{p}{\hbar}$} \, x \, ] + 
j \, \tilde{t} \, \exp[ \, - \,  
\mbox{$\frac{p}{\hbar}$} \, x \,]~.
\end{equation}
In Eqs. (\ref{psii}) and 
(\ref{psit}), we have respectively omitted the 
complex exponential solutions $\exp[  \, - 
\mbox{$\frac{p}{\hbar}$} \, x \, ]$ and 
$\exp[  \, 
\mbox{$\frac{p}{\hbar}$} \, x \, ]$
because they are
in conflict with the boundary condition that $\Psi(x)$ remain finite as
$x \to - \infty$ and $x \to + \infty$. In order to determine the complex 
amplitudes  
$r$, $t$, $\tilde{r}$ and 
$\tilde{t}$, we match the wave function and its slope at the 
discontinuities of the potential (see appendix).  

By using the continuity equation, we can immediately obtain the standard
relation between the transmission and reflection 
coefficients, $t$ and $r$. In fact, Eq.~(\ref{con})  implies that 
the current density 
\[
\mathcal{J} = \mbox{\footnotesize $\frac{p}{2m}$} \, \left\{ \, 
 \left[ \partial_{\x} \overline{\Psi}(x) \right] \, i \, \Psi(x) -
\overline{\Psi}(x) \, i \, 
\partial_{\x} \Psi(x)  \, \right\} 
\]
has the same value at all points $x$. In the free potential regions,
the probability current densities are given by
$
\mathcal{J}_{\mi} = \mbox{\footnotesize $\frac{p}{m}$} \,
\left( \, 1 - |r|^{\2} \, \right)$ and 
$\mathcal{J}_{\pl} = \mbox{\footnotesize $\frac{p}{m}$} \,
|t|^{\2}$. 
Consequently, we find 
\begin{equation} 
|r|^{\2} + |t|^{\2} = 1~.
\end{equation} 
In Figs.\,1 and 2, we plot the transmission probability 
$|t|^{\2}$ as a function of $E[\mbox{eV}]$ for quaternionic potentials of 
different widths and heights (we assume that the incident 
particle is an electron). The presence of a quaternionic perturbation 
potential  modifies the shape of the (complex) transmission probability  
curve. We have a reduction of the transmission probability. 
The presence of an inflection point is evident
by increasing the width and the height of quaternionic potentials. 
Fig.\,2 shows  the transmission probability  
for the quaternionic potential
\[
V + j \, |W| = \left( 2.0 + j \, 1.5 \right) \, \mbox{eV}
\]
and the  complex {\em comparative} barrier~\cite{DAV89}
\[
Z=\sqrt{V^{\2} + |W|^{\2}}=2.5 \, \mbox{eV}~.
\]
The wave numbers for these potentials are\\

\noindent
\begin{tabular}{crcll}
$\bullet$ & $E > \sqrt{V^{\2} + |W|^{\2}}$ & ~:~ & 
~~~$z_{\mi}^{\Z} = i \,  
 \mbox{\footnotesize{$ \sqrt{ \frac{2m}{\hbar^{\2}} \, \left( 
E - \sqrt{V ^{\2}+|W|^{\2}} \right)}$}}$ & $\in i \, \mathbb{R}_{\pl}$\\
 & &  & 
$z_{\mi}^{\V , \W} = i \,  
 \mbox{\footnotesize{$ \sqrt{ \frac{2m}{\hbar^{\2}} \, \left( 
\sqrt{E^{\2} - |W|^{\2}} - V \right)}$}}$ & $\in i \, \mathbb{R}_{\pl}$\\
 &  & & 
$z_{\pl}^{\V , \W} = ~ \,    
 \mbox{\footnotesize{$ \sqrt{ \frac{2m}{\hbar^{\2}} \, \left( 
\sqrt{E^{\2} - |W|^{\2}} + V \right)}$}}$
& $\in ~ \, \mathbb{R}_{\pl}$\\
$\bullet$ & $|W| < E <  \sqrt{V^{\2} + |W|^{\2}}$ & ~:~ & 
~~~$z_{\mi}^{\Z} = ~ \,  
 \mbox{\footnotesize{$ \sqrt{ \frac{2m}{\hbar^{\2}} \, \left( 
\sqrt{V ^{\2}+|W|^{\2}} - E  \right)}$}}$ & $\in ~ \, \mathbb{R}_{\pl}$\\
 & & &
$z_{\pem}^{\V , \W} =  ~ \,  
 \mbox{\footnotesize{$ \sqrt{ \frac{2m}{\hbar^{\2}} \, \left( 
V \pm \sqrt{E^{\2} - |W|^{\2}} \right)}$}}$ & $\in ~ \, \mathbb{R}_{\pl}$\\
$\bullet$ & $ E < |W|~~~~~~~~~~~$ & ~:~ & 
~~~$z_{\mi}^{\Z} = ~ \,  
 \mbox{\footnotesize{$ \sqrt{ \frac{2m}{\hbar^{\2}} \, \left( 
\sqrt{V ^{\2}+|W|^{\2}} - E  \right)}$}}$ & $\in ~ \, \mathbb{R}_{\pl}$\\
 & & & 
$z_{\pem}^{\V , \W} =  ~ \,  \mbox{\footnotesize{$ 
\sqrt{ \frac{2m}{\hbar^{\2}} \, 
\sqrt{V^{\2} +  |W|^{\2} - E^{\2}}
\, \exp \left[
\pm \, i \, \varphi 
\right]}$}}$ & $\in ~ \, \mathbb{C}$\\
& & & 
\end{tabular}

\noindent where $\varphi = \arctan
\mbox{\footnotesize{$\left[ \sqrt{ 
\left( |W|^{\2}-E^{\2} \right) / V^{\2}} \right]$}}$.
For small quaternionic perturbations, the complex {\em comparative} 
barrier $Z$ represents a good approximation of the quaternionic potential 
$V+j \, |W|$. In this case,
\[ Z \sim V \left( 1 + \mbox{$\frac{1}{2}$} \, \frac{|W|^{\2}}{V^{\2}} \right)
~~~\mbox{and}~~~z_{\mi}^{\V , \W} \sim z_{\mi}^{\Z}~.
\]
The anti-self-adjoint operators corresponding to the quaternionic potential 
$V+j \, |W|$ and to the complex barrier
$Z$  are respectively
\[
\mathcal{A}^{\V , \W}_{\E} = i \, \left[ \, 
\mbox{$\frac{\hbar^{\2}}{2m}$} \, \nabla^{\2} 
- V  \, \right]
+ j \, W(\r ,t)~~~\mbox{and}~~~
\mathcal{A}^{\Z}_{\E} = i \, \left[ \, 
\mbox{$\frac{\hbar^{\2}}{2m}$} \, \nabla^{\2} 
- \sqrt{V^{\2}+|W|^{\2}}  \, \right]~.
\]
By using these operators, we can write down two  
complex wave equations
\begin{equation}
\label{ce}
\left[ \mathcal{A}^{\V , \W}_{\E}  \right]^{\2} \, \Psi(x) = 
- E^{\2} \, \Psi (x)~~~\mbox{and}~~~
\left[ \mathcal{A}^{\Z}_{\E}  \right]^{\2} \, \Psi(x) = 
- E^{\2} \, \Psi (x)~.
\end{equation}
The complex operators
\[
\left[ \mathcal{A}^{\V , \W}_{\E}  \right]^{\2} = -
 \left[ \, 
\mbox{$\frac{\hbar^{\2}}{2m}$} \, \nabla^{\2} 
- V  \, \right]^{\2} - |W|^{\2} = 
- \left( \mbox{$\frac{\hbar^{\2}}{2m}$} \right)^{\2} \, \nabla^{\4}
+ 2 \, V \,  \mbox{$\frac{\hbar^{\2}}{2m}$} \, 
\nabla^{\2}  - V^{\2}
-|W|^{\2} 
\]
and
\[
\left[ \mathcal{A}^{\V , \W}_{\E}  \right]^{\2} = -
 \left[ \, 
\mbox{$\frac{\hbar^{\2}}{2m}$} \, \nabla^{\2} 
-  \sqrt{V^{\2}+|W|^{\2}}  \, \right]^{\2} = 
- \left( \mbox{$\frac{\hbar^{\2}}{2m}$} \right)^{\2} \, \nabla^{\4}
+ 2 \,  \sqrt{V^{\2}+|W|^{\2}} \,  
\mbox{$\frac{\hbar^{\2}}{2m}$} \, \nabla^{\2}  - V^{\2}
-|W|^{\2}
\]
can be now easily compared. 
The difference is due to the 
factor which multiplies  $\nabla^{\2}$. Thus, complex 
comparative barriers only represent a first approximation to
quaternionic potentials. In general, we have to consider the pure quaternionic
potential, $j \, |W|$,  as a perturbation effect on the complex barrier 
$V$. 

Deviations from (complex) quantum mechanics appear in proximity of the 
complex barrier $V$  when a quaternionic perturbation is turned on.
Actually, in quaternionic quantum mechanics we find an additional evanescent 
probability of transmission. that is  $|\tilde{t}|^{\2}$. 
This probability as a function of  $E[\mbox{eV}]$ is drawn  in Fig.\,3  
for different values of $x$. 

To conclude  the discussion of quaternionic one-dimensional time invariant 
potentials, we analyze the transmission probability $|t|^{\2}$ as a 
function of the width of complex and quaternionic potentials. In Fig.\,4,
we plot  the transmission probability for critical values of $E$. 
For $E > \sqrt{V^{\2}+|W|^{\2}}$, the minimum value of the transmission 
probability oscillation decreases when the quaternionic perturbation 
increases. In Fig.\,5, we compare complex and {\em pure} quaternionic
potentials covering the same area, $a \, V= b \, |W|$. Clear deviations 
from complex quantum mechanics appear.


\section*{IV. TIME REVERSAL VIOLATING (TRV) POTENTIAL BARRIER}

Let us modify the previous potential barrier by introducing a time reversal 
violating space-dependent phase $\theta(x)$. We shall consider, for the 
region III, the following cases: 
\begin{center}
{\color{blue}{
\begin{tabular}{ccccccccccccccc}
& &  & & & & &  &  & & \\
&   & 
{\sc Region III}$_{\0}$ &  & 
{\sc Region III}$_{\0}$ &   & 
{\color{red}{{\sc Region III}$_{\th}$}} &   & 
{\color{red}{{\sc Region III}$_{\th}$}} &   & 
~~~ \\ 
& &  & & & & &  & \\
 &  & 
{\color{black}{
$\left[ \, V \, ; \, |W| \, , \, {\color{blue}{0}}  \, 
\right]$}}            &  & 
{\color{black}{$\left[ \, V \, ; \, |W| \, , \, {\color{blue}{0}} \, 
\right]$}}            &  & 
{\color{black}{$\left[ \, V \, ; \, |W| \, , \, {\color{red}{\theta}}  \, 
\right]$}}            &  & 
{\color{black}{$\left[ \, V \, ; \, |W| \, , \, {\color{red}{\theta}}  \, 
\right]$}}            &  & 
\\
& &  & & & & &  &  & & \\
  &   & 
{\sc Region III}$_{\0}$ &  & 
{\color{red}{{\sc Region III}$_{\th}$}} &   & 
{\color{red}{{\sc Region III}$_{\th}$}} &   & 
{\sc Region III}$_{\0}$ &   & 
~~~ \\ 
& &  & & & & & & &  & \\
  &  & 
{\color{black}{$\left[ \, V \, ; \, |W| \, , \, {\color{blue}{0}}  \, 
\right]$}}             &  & 
{\color{black}{$\left[ \, V \, ; \, |W| \, , \, {\color{red}{\theta}}  \, 
\right]$}} &  & 
{\color{black}{$\left[ \, V \, ; \, |W| \, , \, {\color{red}{\theta}}  \, 
\right]$}}  &  & 
{\color{black}{$\left[ \, V \, ; \, |W| \, , \, {\color{blue}{0}}  \, 
\right]$}}      &  & 
\\
& &  & & & & & & &  & \\ 
&   & 
{\color{red}{{\sc Region III}$_{\th}$}} &  & 
{\color{red}{{\sc Region III}$_{\th}$}} &   & 
{\sc Region III}$_{\0}$ &   & 
{\sc Region III}$_{\0}$ &   & 
~~~ \\ 
& &  & & & & &  & \\
      &  & 
{\color{black}{$\left[ \, V \, ; \, |W| \, , \, {\color{red}{\theta}}  \, 
\right]$}}            &  & 
{\color{black}{$\left[ \, V \, ; \, |W| \, , \, {\color{red}{\theta}}  \, 
\right]$}}  
 &  & 
{\color{black}{$\left[ \, V \, ; \, |W| \, , \, {\color{blue}{0}}  \, 
\right]$}}            &  & 
{\color{black}{$\left[ \, V \, ; \, |W| \, , \, {\color{blue}{0}} \, 
\right]$}}      
\\
& &  & & & & &  &  & & \\
\hline
& $^{|}$ & 
& $^{|}$ & 
& $^{|}$  & 
& $^{|}$  & 
&  $^{|}$ & \\
& $-b$ & 
& $-c$ &
& $0$ & 
& $+c$  & 
&  $+b$ & \\
\end{tabular}
}}
\end{center}

\noindent As remarked in the introduction, quaternionic deviations from 
complex quantum mechanics could be observed by considering left and right
transmissions through the same quaternionic potential barrier. The left
transmission ($x<-a$) for the quaternionic potential of height $|W|$ and
phase
\begin{equation}
\label{eq1}
\theta(x) = \left\{ \begin{array}{rl} 
0 & ~~~-b<x<0\\
 & \\
\theta & \, ~~~~~~0<x<b
\end{array} \right.
\end{equation}
is obviously equivalent to the right transmission ($x>a$) for the 
quaternionic potential of height $|W|$ and phase
\begin{equation}
\label{eq2}
\theta(x) = \left\{ \begin{array}{rl} 
\theta & ~~~-b<x<0\\
 & \\
0 & \, ~~~~~~0<x<b
\end{array} \right.~.
\end{equation}
By using the transformation (\ref{tran}), we can replace the phase (\ref{eq2}) 
by
\begin{equation}
\label{eq3}
\theta(x) = \left\{ \begin{array}{rl} 
0 & ~~~-b<x<0\\
 & \\
- \theta & \, ~~~~~~0<x<b
\end{array} \right.
\end{equation}
Thus, the plot of the transmission coefficient as a function of $\theta[\pi]$
is a valid indicator of possible deviations from complex quantum mechanics.
Symmetric curves (around the point $\theta[\pi]=1$) shall imply {\em no}
difference between left and right transmission through the same quaternionic 
barrier. In Figs.\, 6, 7 and 8, we draw the transmission probability, 
$|t|^{\2}$, and the absolute value of the transmission coefficient, $|Arg(t)|$,
as a function of the phase $\theta[\pi]$. Qualitative deviations for 
complex quantum mechanics appear for {\em asymmetric} time violating 
potentials. It is also interesting to note that by increasing the phase
($\theta[\pi] \to 1$), quaternionic perturbation effects 
are {\em minimized}. For the convenience of the reader
we explicitly give (see  Tables 1, 2 and 3) the transmission 
probability $|t|^{\2}$ and the 
transmission coefficient $t$ for different values of the potential phase
$\theta$ and the electron energy $E$.


\section*{VI. CONCLUSIONS}

Very little progress in mathematical 
understanding of quaternionic analysis and algebra 
have often created (and sometimes justified) a distrust feeling to
quaternionic formulations of physical theories.  From our point of view,
the recent  increasing  improvement  of the mathematical structures involved 
in the quaternionic quantum mechanics could result in a rapid progress in this
subject.

The usefulness of quaternions (and, more in general, Clifford algebras) to 
unify algebraic and geometric aspects in  discussing special relativity,
Maxwell and Dirac equations  is universally recognized.  
Nevertheless, notwithstanding the substantial literature analyzing 
quaternionic physical theories, a strong motivation forcing the use 
of quaternions {\em instead} of complex numbers is lacking.
The experimental proposals of Peres~\cite{PER79}, the theoretical analysis
of Davies and McKellar~\cite{DAV89,DAV92} and the 
detailed and systematic development of quaternionic quantum mechanics 
in the Adler's 
book~\cite{ADL} surely represent the milestone in  looking for quaternionic 
deviations from complex quantum mechanics. 

In this paper, we have presented
a complete phenomenology of the quaternionic potential barrier by discussing
the time invariant and time violating case. Interesting features of 
quaternionic perturbation effects emerge in the transmission and 
reflection coefficients. The various graphs  show how the quantum  
measurement theory may be affected by 
changing from complex to quaternionic systems. 
The present work represents a preliminary step towards a
significant advance in understanding quaternionic potentials and 
in looking for their experimental evidence. An interesting discussion
about quaternionic violations of the algebraic relationship between the 
six coherent cross sections of any three scatterers, taken singly and 
pairwise, is found in~\cite{PER96}.

Quaternionic time violating
potentials and quaternionic perturbations (wich minimize the deviations
from complex quantum mechanics) could play an important role in the 
CP violating physics. 
A theoretical discussion based on the wave packet formalism will be 
necessary to analyse experimental tests based on kaon
regeneration~\cite{PER79,PER96}. For asymmetric potentials a  
non null signals of quaternionic (time violating)  effects should be 
observed.  We will try to develop the wave packet treatment in a later 
article.


\subsection*{Acknowledgements}

This work was started during the stay of S.\,D.\.L and G.\.C.\,D. at the
Department of Physics, University of Lecce. 
The authors acknowledge INFN, CAPES and FAEP for financial support
and are grateful to S.~Marchiafava and  P.~Rotelli for helpful comments 
and suggestions.



\newpage

\section*{APPENDIX. MACTHING CONDITIONS}

\subsection*{A. TRI POTENTIAL BARRIER}

The matching conditions for the TRI potential barrier imply 
\begin{equation}
\label{rt}
\begin{array}{ccl}
\left( 
\begin{array}{c}
1 \\
r \\
\tilde{r} \\
\tilde{r} 
\end{array}
\right)
& = &
\mathcal{S}  \, [a,b; E; V, |W|]
\, \left( 
\begin{array}{c}
t \\
t \\
\tilde{t} \\
\tilde{t}
\end{array}
\right)~,
\end{array}
\end{equation}
where
\begin{eqnarray*}
\mathcal{S}  \, [a,b;E;V, |W|]  & = &
\underbrace{D_{\mi}
\,
A_{\mi}}_{\mbox{S[I$_{\mi}$]}} \,  
 \, \underbrace{M^{\V} 
D_{\b \mi \a}^{\V} \, [ M^{\V} ]^{\mi \1}}_{{\color{blue}{\mbox{S[II]}}}} \,  
\times \\
 & & 
\underbrace{Q^{\Wm} \, M^{\V , \Wm} \,
D_{\mi \2 \b}^{\V , \Wm} \, 
[ M^{\V , \Wm} ]^{\mi \1} \, [ Q^{\Wm}]^{\mi \1}}_{
{\color{red}{\mbox{S[III]}}}} \,  \times \\
& & \underbrace{M^{\V} \, 
D_{\b \mi \a}^{\V} \,
[ M^{\V} ]^{\mi \1}}_{{\color{blue}{\mbox{S[II]}}}} \,  \,
\underbrace{A_{\pl}
\, D_{\pl}}_{\mbox{S[I$_{\pl}$]}} 
\end{eqnarray*}
and
\[
\begin{array}{lcl}
D_{\mi} & = & \mbox{diag} \left\{ \, 
\exp \left[ \, i  \, 
\mbox{$\frac{p}{\hbar}$} \, a \, \right] \, , \, 
\exp \left[ \, - \, i  \, 
\mbox{$\frac{p}{\hbar}$} \, a \, \right] \, , \, 
\exp \left[ \, 
\mbox{$\frac{p}{\hbar}$} \, a \, \right] \, , \, 
\exp \left[ \, 
\mbox{$\frac{p}{\hbar}$} \, a \, \right]
\, \right\}~,\\ [4mm]
A_{\mi} & ~=~ &  
\mbox{\footnotesize $\frac{1}{2}$} \,
\left(
\begin{array}{rr}
 1 \, &  $-$ \,  i  \, 
\mbox{$\frac{\hbar}{p}$} \\
1 \,  & 
 i  \, 
\mbox{$\frac{\hbar}{p}$} 
\end{array}
\right)
\, \oplus \,
\left( 
\begin{array}{cc}
1 &
0 \\
0 & 
\mbox{$\frac{\hbar}{p}$} 
\end{array}
\right) 
~,\\ [4mm]
M^{\V , \Wm} & ~=~  & 
\left( 
\begin{array}{cc}
1 & 1  \\
z_{\mi}^{\E; \V , \Wm} & $-$ z_{\mi}^{\E; \V , \Wm} 
\end{array}
\right)
\, \oplus  \, 
\left( 
\begin{array}{cc}
1 & 1\\
z_{\pl}^{\E ; \V , \Wm} & $-$ z_{\pl}^{\E ; \V , \Wm}
\end{array}
\right)~,\\  [6mm]
M^{\V } & = & M^{\V , \Wm \tot \0} ~,\\ [4mm]  
Q^{\Wm,\te}  & ~=~ & 
\left( 
\begin{array}{cc}
1 & [ v^{\E;\Wm , \te} ]_{\Co}  \\
{[} $-$ j  u^{\E;\Wm , \te} {]}_{\Co}  & 1
\end{array}
\right) 
\, \otimes  \, 
\left(
\begin{array}{cc}
1 & 0  \\
0 & 1
\end{array}
\right)~,\\  [6mm]
Q^{\Wm } & = &  Q^{\Wm , \te \tot \0} ~,\\ [4mm] 
D_{\e}^{\V, \Wm} & = & \mbox{diag} \left\{ \,
\exp \left[ \, z_{\mi}^{\E ; \V, \Wm} \, \eta \, \right]
\, , \, 
\exp \left[ \, - \, z_{\mi}^{\E ; \V , \Wm} \, \eta \, \right]
\, , \, 
\exp \left[ \, z_{\pl}^{\E ; \V , \Wm} \, \eta \, \right]
\, , \, 
\exp \left[ \, - \, z_{\pl}^{\E; \V , \Wm} \, \eta \, \right]
\, \right\}~,\\  [4mm]
D_{\e}^{\V }  & = &   D_{\e}^{\V , \Wm \tot \0} ~,
\\ [4mm]
A_{\pl} & ~=~ &
\left(
\begin{array}{cc}
1 & 0 \\
0  & 
 i  \, 
\mbox{$\frac{p}{\hbar}$} 
\end{array}
\right)
\, \oplus  \, 
\left( 
\begin{array}{cc}
1 &
0 \\
0 & 
$-$ \, \mbox{$\frac{p}{\hbar}$} 
\end{array}
\right)~,\\[6mm]
D_{\pl} & ~=~ & \mbox{diag} \left\{ \, 
\exp \left[ \, i  \, 
\mbox{$\frac{p}{\hbar}$} \, a \, \right]
\, , \, 
\exp \left[ \, i  \, 
\mbox{$\frac{p}{\hbar}$} \, a \, \right]
\, , \, 
\exp \left[ \, - \, 
\mbox{$\frac{p}{\hbar}$} \, a \, \right]
\, , \, 
\exp \left[ \, - \, 
\mbox{$\frac{p}{\hbar}$} \, a \, \right]
\, \right\}~.
\end{array}
\]

\noindent
The {\em complex limit} is obtained by setting $b=0$.  
In this case ($\mbox{S[III]} = \boldsymbol{1}$) $\mathcal{S}  \, [a,b;E; V,W]$
reduces to
\[
\mathcal{S}  \, [a;E;V] =
D_{\mi}
\,
A_{\mi} \, 
M^{\V} \,
D_{\mi \2 \a}^{\V} \, 
[ M^{\V}]^{\mi \1} \,
A_{\pl}
\, D_{\pl}~.  
\]
By matrix algebra, we easily calculate the coefficients for reflection and 
tran\-smis\-sion 
\[
\begin{array}{ccl}
t & = & 
 \mbox{\footnotesize $\exp \left[ \, - \, 2 \, i \, 
 \mbox{\tiny $
  \frac{p}{\hbar}$} \, a \, \right]$} \, 
\left\{ \, 
\cosh \left[ \, 2 \,  z_{\mi}^{\E;\V} \, a \, \right]
+ \,  \mbox{\footnotesize $
  \frac{i}{2}$} \, \chi_{\mi} \, 
\sinh \left[ \, 2 \,  z_{\mi}^{\E;\V} \, a \, \right] \,
\right\}^{\mi \1}~,\\
r & = & - \,   \mbox{\footnotesize $
  \frac{i}{2}$} \, \chi_{\pl} \, 
\sinh \left[ \, 2 \, z_{\mi}^{\E;\V}
\, a \, \right] \, t\\
\tilde{t} & = & 0~,\\
\tilde{r} & = & 0~,
\end{array}
\]
where $\chi_{\pem} = \frac{\hbar}{p} \, z_{\mi}^{\E;\V} \pm 
\left( \frac{\hbar}{p} \, z_{\mi}^{\E;\V} \right)^{\mi \1}$.

\newpage

\subsection*{B. TRV POTENTIAL BARRIER}

The matrix $\mathcal{S}  \, [a,b,c; E; V, |W|,\theta]$ is now expressed in 
terms of
\[
\mbox{S[III]} =
\left\{ \begin{array}{ccl}
\mbox{S[III}_{{\color{blue}{\0 \, \0}} \, 
{\color{red}{\th \, \th}}}] & : & S[0,-b] \, \times \, 
S[\theta ,-b]\\
 & & \\
\mbox{S[III}_{{\color{blue}{\0}} \, {\color{red}{\th \, \th}} \, 
{\color{blue}{\0}}}] & : &  
S[0,c-b] \, \times \, 
S[\theta ,-2c] \, \times \, S[0,c-b]\\
 & & \\
\mbox{S[III}_{{\color{red}{\th \, \th}} \, {\color{blue}{\0 \, \0}}}] & : & 
S[\theta,-b] \, \times \, 
S[0 ,-b]\\
\end{array}
\right.
\] 
where
\begin{eqnarray*}
S[\theta,\eta] & = &    
Q^{\Wm,\te} \, M^{\V , \Wm} \,
D_{\e}^{\V , \Wm} \, 
[ M^{\V , \Wm} ]^{\mi \1} \, [ Q^{\Wm,\te}]^{\mi \1}~.
\end{eqnarray*}

\newpage

\section*{Table 1 [Fig. 6]} 

\vspace*{0.5cm}

\begin{center}
$E=0.5 \, \, \mbox{eV}$~;~~
${\color{blue}{V}}={\color{red}{2 \, |W|}}= 2 \, \, \mbox{eV}$~;
~~${\color{blue}{a}}= {\color{red}{2 \, b}}= {\color{red}{4 \, c}}
= 1 \, \overset{\circ}{\mbox{A}}$
\end{center}

\vspace*{1cm}

\begin{center}
\begin{tabular}{|c||c|c|c|}\hline
 & & & \\
$\left[ \, E \, ; \, {\color{blue}{V}} \, , \, {\color{red}{|W|}} \, \right]$ &
${\color{red}{\theta}}$ &
$|t|^{\2}$ &
$ t $ \\
 & & & \\ \hline
 & & & \\
$\left[ \, 0.5 \, ; \, {\color{blue}{2}} \, , \, {\color{red}{0}} \, \right]$ &
$\left[ \, 0 \, , \, 0 \, ; \, 0 \, , \, 0 \, \right]$
 & ~$0.62596$~ & ~$0.618647 - i \, 0.493189$~\\ 
 & & & \\ \hline \hline
 & & & \\
$\left[ \, 0.5 \, ; \, {\color{blue}{2}} \, , \, {\color{red}{1}} \, \right]$ &
{\color{blue}{$\left[ \, 0 \, , \, 0 \, ; \, 0 \, , \, 0 \, \right]$}} & 
~{\color{blue}{$0.612889$}}~ & ~{\color{blue}{$0.605527 - i \, 0.496212$}}~\\ 
 & & & \\ \hline \hline
 & & & \\
 & $\left[ \, 0 \, , \, 0  \, ; 
\, {\color{red}{\frac{\pi}{6}}}  \, , \, {\color{red}{\frac{\pi}{6}}} 
\, \right]$ & ~{\color{blue}{$0.613720$}}~ & ~$0.606559 - i \, 0.495789$~\\ 
$\left[ \, 0.5 \, ; \, {\color{blue}{2}} \, , \, {\color{red}{1}} \, \right]$
 & $\left[ \, 0 \, , \, {\color{red}{\frac{\pi}{6}}}  \, ; 
\, {\color{red}{\frac{\pi}{6}}} \,, \,  0  \, \right]$ & 
~$0.613748$~ & ~$0.606389 - i \, 0.496025$~\\ 
 & $\left[ \, {\color{red}{\frac{\pi}{6}}}  \, , 
\, {\color{red}{\frac{\pi}{6}}} \, ; \, 0 \, , \,  0 \, \right]$ & 
~{\color{blue}{$0.613720$}}~ & ~$0.606159 - i \, 0.496278$~\\
 & & & \\ \hline
 & & & \\
 & $\left[ \, 0 \, , \, 0  \, ; 
\, {\color{red}{\frac{\pi}{4}}}  \, , \, {\color{red}{\frac{\pi}{4}}} 
\, \right]$ & 
~{\color{blue}{$0.614708$}}~ & ~$0.60763 - i \, 0.495473$~\\ 
$\left[ \, 0.5 \, ; \, {\color{blue}{2}} \, , \, {\color{red}{1}} \, \right]$
 & $\left[ \, 0 \, , \, {\color{red}{\frac{\pi}{4}}}  \, ; 
\, {\color{red}{\frac{\pi}{4}}} \,, \,  0  \, \right]$ & 
~$0.614769$~ & ~$0.607413 - i \, 0.495800$~\\ 
 & $\left[ \, {\color{red}{\frac{\pi}{4}}}  \, , 
\, {\color{red}{\frac{\pi}{4}}} \, ; \, 0 \, , \,  0 \, \right]$ & 
~{\color{blue}{$0.614708$}}~ & ~$0.607065 - i \, 0.496165$~\\
 & & & \\ \hline
 & & & \\
 & $\left[ \, 0 \, , \, 0  \, ; 
\, {\color{red}{\frac{\pi}{3}}}  \, , \, {\color{red}{\frac{\pi}{3}}} 
\, \right]$ & 
~{\color{blue}{$0.615996$}}~ & ~$0.608983 - i \, 0.495113$~\\ 
$\left[ \, 0.5 \, ; \, {\color{blue}{2}} \, , \, {\color{red}{1}} \, \right]$
 & $\left[ \, 0 \, , \, {\color{red}{\frac{\pi}{3}}}  \, ; 
\, {\color{red}{\frac{\pi}{3}}} \,, \,  0  \, \right]$ & 
~$0.616100$~ & ~$0.608749 - i \, 0.495504$~\\ 
 & $\left[ \, {\color{red}{\frac{\pi}{3}}}  \, , 
\, {\color{red}{\frac{\pi}{3}}} \, ; \, 0 \, , \,  0 \, \right]$ & 
~{\color{blue}{$0.615996$}}~ & ~$0.608291 - i \, 0.495962$~\\
 & & & \\ \hline
 & & & \\
 & $\left[ \, 0 \, , \, 0  \, ; 
\, {\color{red}{\frac{\pi}{2}}}  \, , \, {\color{red}{\frac{\pi}{2}}} 
\, \right]$ & 
~{\color{blue}{$0.619112$}}~ & ~$0.612155 - i \, 0.494347$~\\ 
$\left[ \, 0.5 \, ; \, {\color{blue}{2}} \, , \, {\color{red}{1}} \, \right]$
 & $\left[ \, 0 \, , \, {\color{red}{\frac{\pi}{2}}}  \, ; 
\, {\color{red}{\frac{\pi}{2}}} \,, \,  0  \, \right]$ & 
~$0.619321$~ & ~$0.611982 - i \, 0.494772$~\\ 
 & $\left[ \, {\color{red}{\frac{\pi}{2}}}  \, , 
\, {\color{red}{\frac{\pi}{2}}} \, ; \, 0 \, , \,  0 \, \right]$ & 
~{\color{blue}{$0.619112$}}~ & ~$0.611358 - i \, 0.495332$~\\
 & & & \\ \hline
& & & \\
 & {\color{blue}{$\left[ \, 0 \, , \, 0  \, ; 
\, \pi  \, , \, \pi 
\, \right]$}} & 
~{\color{blue}{$0.625369$}}~ & ~{\color{blue}{$0.618021 - i \, 0.493376$}}~\\ 
$\left[ \, 0.5 \, ; \, {\color{blue}{2}} \, , \, {\color{red}{1}} \, \right]$
 & $\left[ \, 0 \, , \, \pi  \, ; 
\, \pi \,, \,  0  \, \right]$ & 
~$0.625792$~ & ~$0.618478 - i \, 0.493232$~\\
 & {\color{blue}{$\left[ \, \pi  \, , 
\, \pi \, ; \, 0 \, , \,  0 \, \right]$}} & 
~{\color{blue}{$0.625369$}}~ & ~{\color{blue}{$0.618021 - i \, 0.493376$}}~\\
 & & & \\ \hline
\end{tabular}
\end{center}

\newpage

\section*{Table 2 [Fig. 7]}

\vspace*{0.5cm}

\begin{center}
$E=1.5 \, \, \mbox{eV}$~;~~
${\color{blue}{V}}={\color{red}{2 \, |W|}}= 2 \, \, \mbox{eV}$~;
~~${\color{blue}{a}}= {\color{red}{2 \, b}}= {\color{red}{4 \, c}}
= 1 \, \overset{\circ}{\mbox{A}}$
\end{center}

\vspace*{1cm}

\begin{center}
\begin{tabular}{|c||c|c|c|}\hline
 & & & \\
$\left[ \, E \, ; \, {\color{blue}{V}} \, , \, {\color{red}{|W|}} \, \right]$ &
${\color{red}{\theta}}$ &
$|t|^{\2}$ &
$ t $ \\
 & & & \\ \hline
 & & & \\
$\left[ \, 1.5 \, ; \, {\color{blue}{2}} \, , \, {\color{red}{0}} \, \right]$ &
$\left[ \, 0 \, , \, 0 \, ; \, 0 \, , \, 0 \, \right]$
 & ~$0.845474$~ & ~$0.835936 - i \, 0.382994$~\\ 
 & & & \\ \hline \hline
 & & & \\
$\left[ \, 1.5 \, ; \, {\color{blue}{2}} \, , \, {\color{red}{1}} \, \right]$ &
{\color{blue}{$\left[ \, 0 \, , \, 0 \, ; \, 0 \, , \, 0 \, \right]$}} & 
~{\color{blue}{$0.840310$}}~ & ~{\color{blue}{$0.830647 - i \, 0.387733$}}~\\ 
 & & & \\ \hline \hline
 & & & \\
 & $\left[ \, 0 \, , \, 0  \, ; 
\, {\color{red}{\frac{\pi}{6}}}  \, , \, {\color{red}{\frac{\pi}{6}}} 
\, \right]$ & ~{\color{blue}{$0.840637$}}~ & ~$0.831080 - i \, 0.387223$~\\ 
$\left[ \, 1.5 \, ; \, {\color{blue}{2}} \, , \, {\color{red}{1}} \, \right]$
 & $\left[ \, 0 \, , \, {\color{red}{\frac{\pi}{6}}}  \, ; 
\, {\color{red}{\frac{\pi}{6}}} \,, \,  0  \, \right]$ & 
~$0.840651$~ & ~$0.830996 - i \, 0.387424$~\\ 
 & $\left[ \, {\color{red}{\frac{\pi}{6}}}  \, , 
\, {\color{red}{\frac{\pi}{6}}} \, ; \, 0 \, , \,  0 \, \right]$ & 
~{\color{blue}{$0.840637$}}~ & ~$0.830877 - i \, 0.387660$~\\
 & & & \\ \hline
 & & & \\
 & $\left[ \, 0 \, , \, 0  \, ; 
\, {\color{red}{\frac{\pi}{4}}}  \, , \, {\color{red}{\frac{\pi}{4}}} 
\, \right]$ & 
~{\color{blue}{$0.841023$}}~ & ~$0.831516 - i \, 0.386787$~\\ 
$\left[ \, 1.5 \, ; \, {\color{blue}{2}} \, , \, {\color{red}{1}} \, \right]$
 & $\left[ \, 0 \, , \, {\color{red}{\frac{\pi}{4}}}  \, ; 
\, {\color{red}{\frac{\pi}{4}}} \,, \,  0  \, \right]$ & 
~$0.841055$~ & ~$0.831409 - i \, 0.387058$~\\ 
 & $\left[ \, {\color{red}{\frac{\pi}{4}}}  \, , 
\, {\color{red}{\frac{\pi}{4}}} \, ; \, 0 \, , \,  0 \, \right]$ & 
~{\color{blue}{$0.841023$}}~ & ~$0.831228 - i \, 0.387406$~\\
 & & & \\ \hline
 & & & \\
 & $\left[ \, 0 \, , \, 0  \, ; 
\, {\color{red}{\frac{\pi}{3}}}  \, , \, {\color{red}{\frac{\pi}{3}}} 
\, \right]$ & 
~{\color{blue}{$0.841526$}}~ & ~$0.832061 - i \, 0.386266$~\\ 
$\left[ \, 1.5 \, ; \, {\color{blue}{2}} \, , \, {\color{red}{1}} \, \right]$
 & $\left[ \, 0 \, , \, {\color{red}{\frac{\pi}{3}}}  \, ; 
\, {\color{red}{\frac{\pi}{3}}} \,, \,  0  \, \right]$ & 
~$0.841581$~ & ~$0.831948 - i \, 0.386580$~\\ 
 & $\left[ \, {\color{red}{\frac{\pi}{3}}}  \, , 
\, {\color{red}{\frac{\pi}{3}}} \, ; \, 0 \, , \,  0 \, \right]$ & 
~{\color{blue}{$0.841526$}}~ & ~$0.831708 - i \, 0.387024$~\\
 & & & \\ \hline
 & & & \\
 & $\left[ \, 0 \, , \, 0  \, ; 
\, {\color{red}{\frac{\pi}{2}}}  \, , \, {\color{red}{\frac{\pi}{2}}} 
\, \right]$ & 
~{\color{blue}{$0.842740$}}~ & ~$0.833323 - i \, 0.385115$~\\ 
$\left[ \, 1.5 \, ; \, {\color{blue}{2}} \, , \, {\color{red}{1}} \, \right]$
 & $\left[ \, 0 \, , \, {\color{red}{\frac{\pi}{2}}}  \, ; 
\, {\color{red}{\frac{\pi}{2}}} \,, \,  0  \, \right]$ & 
~$0.842850$~ & ~$0.833247 - i \, 0.385421$~\\ 
 & $\left[ \, {\color{red}{\frac{\pi}{2}}}  \, , 
\, {\color{red}{\frac{\pi}{2}}} \, ; \, 0 \, , \,  0 \, \right]$ & 
~{\color{blue}{$0.842740$}}~ & ~$0.832917 - i \, 0.385991$~\\
 & & & \\ \hline
& & & \\
 & {\color{blue}{$\left[ \, 0 \, , \, 0  \, ; 
\, \pi  \, , \, \pi 
\, \right]$}} & 
~{\color{blue}{$0.845158$}}~ & ~{\color{blue}{$0.835583 - i \, 0.383353$}}~\\ 
$\left[ \, 1.5 \, ; \, {\color{blue}{2}} \, , \, {\color{red}{1}} \, \right]$
 & $\left[ \, 0 \, , \, \pi  \, ; 
\, \pi \,, \,  0  \, \right]$ & 
~$0.845378$~ & ~$0.835836 - i \, 0.383087$~\\
 & {\color{blue}{$\left[ \, \pi  \, , 
\, \pi \, ; \, 0 \, , \,  0 \, \right]$}} & 
~{\color{blue}{$0.845158$}}~ & ~{\color{blue}{$0.835583 - i \, 0.383353$}}~\\
 & & & \\ \hline
\end{tabular}
\end{center}

\newpage

\section*{Table 3 [Fig.8]}

\vspace*{0.5cm}

\begin{center}
$E=3 \, \, \mbox{eV}$~;~~
${\color{blue}{V}}={\color{red}{2 \, |W|}}= 2 \, \, \mbox{eV}$~;
~~${\color{blue}{a}}= {\color{red}{2 \, b}}= {\color{red}{4 \, c}}
= 1 \, \overset{\circ}{\mbox{A}}$
\end{center}

\vspace*{1cm}

\begin{center}
\begin{tabular}{|c||c|c|c|}\hline
 & & & \\
$\left[ \, E \, ; \, {\color{blue}{V}} \, , \, {\color{red}{|W|}} \, \right]$ &
${\color{red}{\theta}}$ &
$|t|^{\2}$ &
$ t $ \\
 & & & \\ \hline
 & & & \\
$\left[ \, 3 \, ; \, {\color{blue}{2}} \, , \, {\color{red}{0}} \, \right]$ &
$\left[ \, 0 \, , \, 0 \, ; \, 0 \, , \, 0 \, \right]$
 & ~$0.925842$~ & ~$0.915930 - i \, 0.294811$~\\ 
 & & & \\ \hline \hline
 & & & \\
$\left[ \, 3 \,; \, {\color{blue}{2}} \, , \, {\color{red}{1}} \, \right]$ &
{\color{blue}{$\left[ \, 0 \, , \, 0 \, ; \, 0 \, , \, 0 \, \right]$}} & 
~{\color{blue}{$0.923710$}}~ & ~{\color{blue}{$0.913674 - i \, 0.298177$}}~\\ 
 & & & \\ \hline \hline
 & & & \\
 & $\left[ \, 0 \, , \, 0  \, ; 
\, {\color{red}{\frac{\pi}{6}}}  \, , \, {\color{red}{\frac{\pi}{6}}} 
\, \right]$ & ~{\color{blue}{$0.923843$}}~ & ~$0.913869 - i \, 0.297802$~\\ 
$\left[ \, 3 \,; \, {\color{blue}{2}} \, , \, {\color{red}{1}} \, \right]$
 & $\left[ \, 0 \, , \, {\color{red}{\frac{\pi}{6}}}  \, ; 
\, {\color{red}{\frac{\pi}{6}}} \,, \,  0  \, \right]$ & 
~$0.923850$~ & ~$0.913822 - i \, 0.297959$~\\ 
 & $\left[ \, {\color{red}{\frac{\pi}{6}}}  \, , 
\, {\color{red}{\frac{\pi}{6}}} \, ; \, 0 \, , \,  0 \, \right]$ & 
~{\color{blue}{$0.923843$}}~ & ~$0.913757 - i \, 0.298147$~\\
 & & & \\ \hline
 & & & \\
 & $\left[ \, 0 \, , \, 0  \, ; 
\, {\color{red}{\frac{\pi}{4}}}  \, , \, {\color{red}{\frac{\pi}{4}}} 
\, \right]$ & 
~{\color{blue}{$0.924000$}}~ & ~$0.914057 - i \, 0.297490$~\\ 
$\left[ \, 3 \,; \, {\color{blue}{2}} \, , \, {\color{red}{1}} \, \right]$
 & $\left[ \, 0 \, , \, {\color{red}{\frac{\pi}{4}}}  \, ; 
\, {\color{red}{\frac{\pi}{4}}} \,, \,  0  \, \right]$ & 
~$0.924016$~ & ~$0.913998 - i \, 0.297699$~\\ 
 & $\left[ \, {\color{red}{\frac{\pi}{4}}}  \, , 
\, {\color{red}{\frac{\pi}{4}}} \, ; \, 0 \, , \,  0 \, \right]$ & 
~{\color{blue}{$0.924000$}}~ & ~$0.913898 - i \, 0.297977$~\\
 & & & \\ \hline
 & & & \\
 & $\left[ \, 0 \, , \, 0  \, ; 
\, {\color{red}{\frac{\pi}{3}}}  \, , \, {\color{red}{\frac{\pi}{3}}} 
\, \right]$ & 
~{\color{blue}{$0.924205$}}~ & ~$0.914289 - i \, 0.297122$~\\ 
$\left[ \, 3 \,; \, {\color{blue}{2}} \, , \, {\color{red}{1}} \, \right]$
 & $\left[ \, 0 \, , \, {\color{red}{\frac{\pi}{3}}}  \, ; 
\, {\color{red}{\frac{\pi}{3}}} \,, \,  0  \, \right]$ & 
~$0.924232$~ & ~$0.914226 - i \, 0.297360$~\\ 
 & $\left[ \, {\color{red}{\frac{\pi}{3}}}  \, , 
\, {\color{red}{\frac{\pi}{3}}} \, ; \, 0 \, , \,  0 \, \right]$ & 
~{\color{blue}{$0.924205$}}~ & ~$0.914095 - i \, 0.297718$~\\
 & & & \\ \hline
 & & & \\
 & $\left[ \, 0 \, , \, 0  \, ; 
\, {\color{red}{\frac{\pi}{2}}}  \, , \, {\color{red}{\frac{\pi}{2}}} 
\, \right]$ & 
~{\color{blue}{$0.924699$}}~ & ~$0.914820 - i \, 0.296317$~\\ 
$\left[ \, 3 \,; \, {\color{blue}{2}} \, , \, {\color{red}{1}} \, \right]$
 & $\left[ \, 0 \, , \, {\color{red}{\frac{\pi}{2}}}  \, ; 
\, {\color{red}{\frac{\pi}{2}}} \,, \,  0  \, \right]$ & 
~$0.924753$~ & ~$0.914776 - i \, 0.296542$~\\ 
 & $\left[ \, {\color{red}{\frac{\pi}{2}}}  \, , 
\, {\color{red}{\frac{\pi}{2}}} \, ; \, 0 \, , \,  0 \, \right]$ & 
~{\color{blue}{$0.924699$}}~ & ~$0.914596 - i \, 0.297006$~\\
 & & & \\ \hline
& & & \\
 & {\color{blue}{$\left[ \, 0 \, , \, 0  \, ; 
\, \pi  \, , \, \pi 
\, \right]$}} & 
~{\color{blue}{$0.925681$}}~ & ~{\color{blue}{$0.915736 - i \, 0.295142$}}~\\ 
$\left[ \, 3 \,; \, {\color{blue}{2}} \, , \, {\color{red}{1}} \, \right]$
 & $\left[ \, 0 \, , \, \pi  \, ; 
\, \pi \,, \,  0  \, \right]$ & 
~$0.925789$~ & ~$0.915873 - i \, 0.294900$~\\
 & {\color{blue}{$\left[ \, \pi  \, , 
\, \pi \, ; \, 0 \, , \,  0 \, \right]$}} &
 ~{\color{blue}{$0.925681$}}~ & ~{\color{blue}{$0.915736 - i \, 0.295142$}}~\\
 & & & \\ \hline
\end{tabular}
\end{center}

\newpage

\begin{figure}
\hspace*{-3.8cm}
\includegraphics[width=10.5cm, height=19.5cm, angle=90]{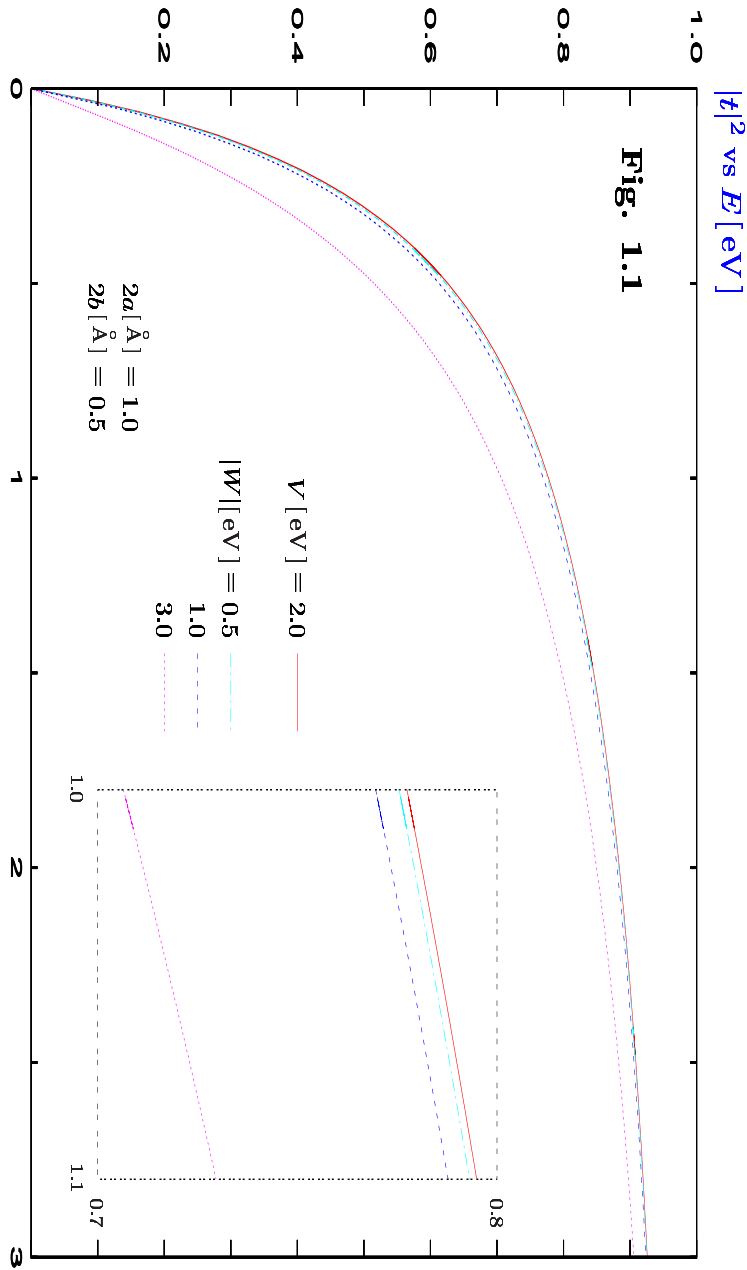}
\hspace*{-3.8cm}
\includegraphics[width=10.5cm, height=19.5cm, angle=90]{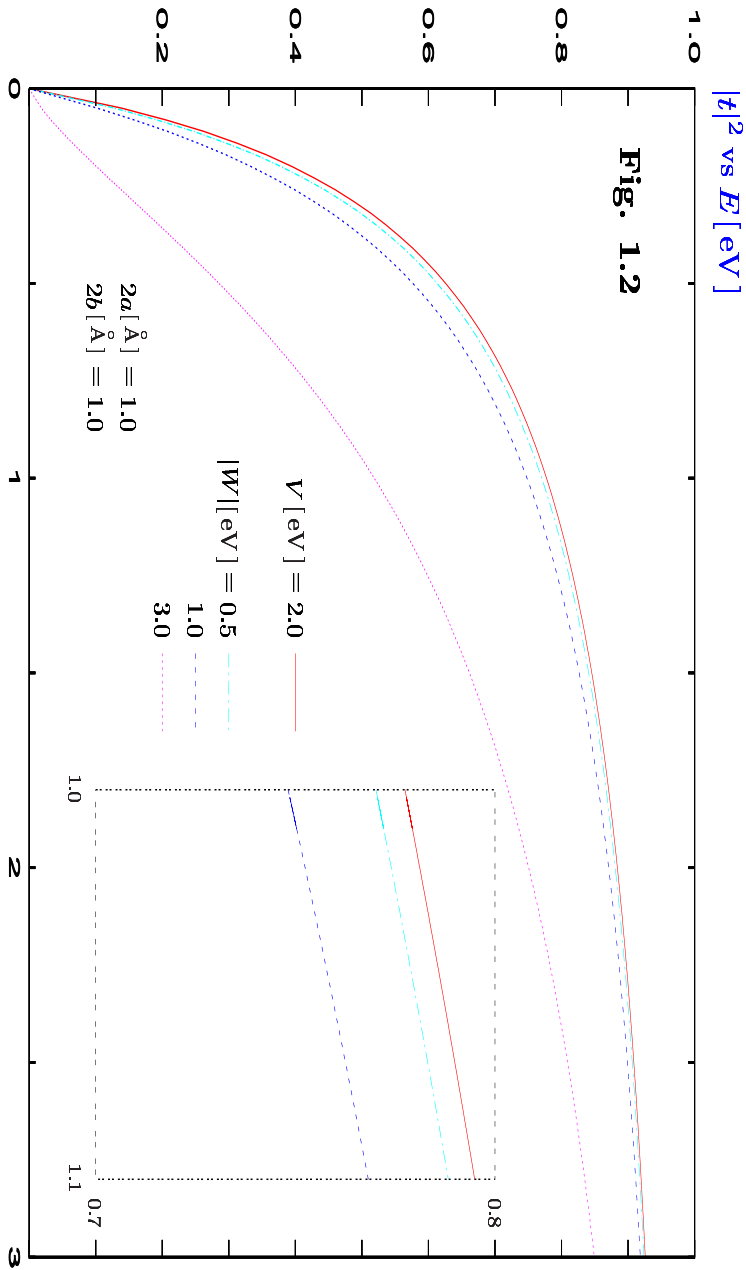}
\caption{Electron transmission probability, $|t|^{\2}$, as a function of 
$E$[eV] for quaternionic time reversal invariant potentials~\cite{DATA}.
The full line indicates the complex quantum 
mechanics result for the potential barrier of width 
$2a$[{\tiny{$\overset{\circ}{A}$}}]$=1.0$ 
and height $V$[eV]$=2.0$. 
The dashed lines (drawn for different values of the width $2b$ and the 
height $|W|$ of the potential $j \, W$)
show the quaternionic perturbation effects.}
\end{figure}

\newpage

\begin{figure}
\hspace*{-3.8cm}
\includegraphics[width=10.5cm, height=19.5cm, angle=90]{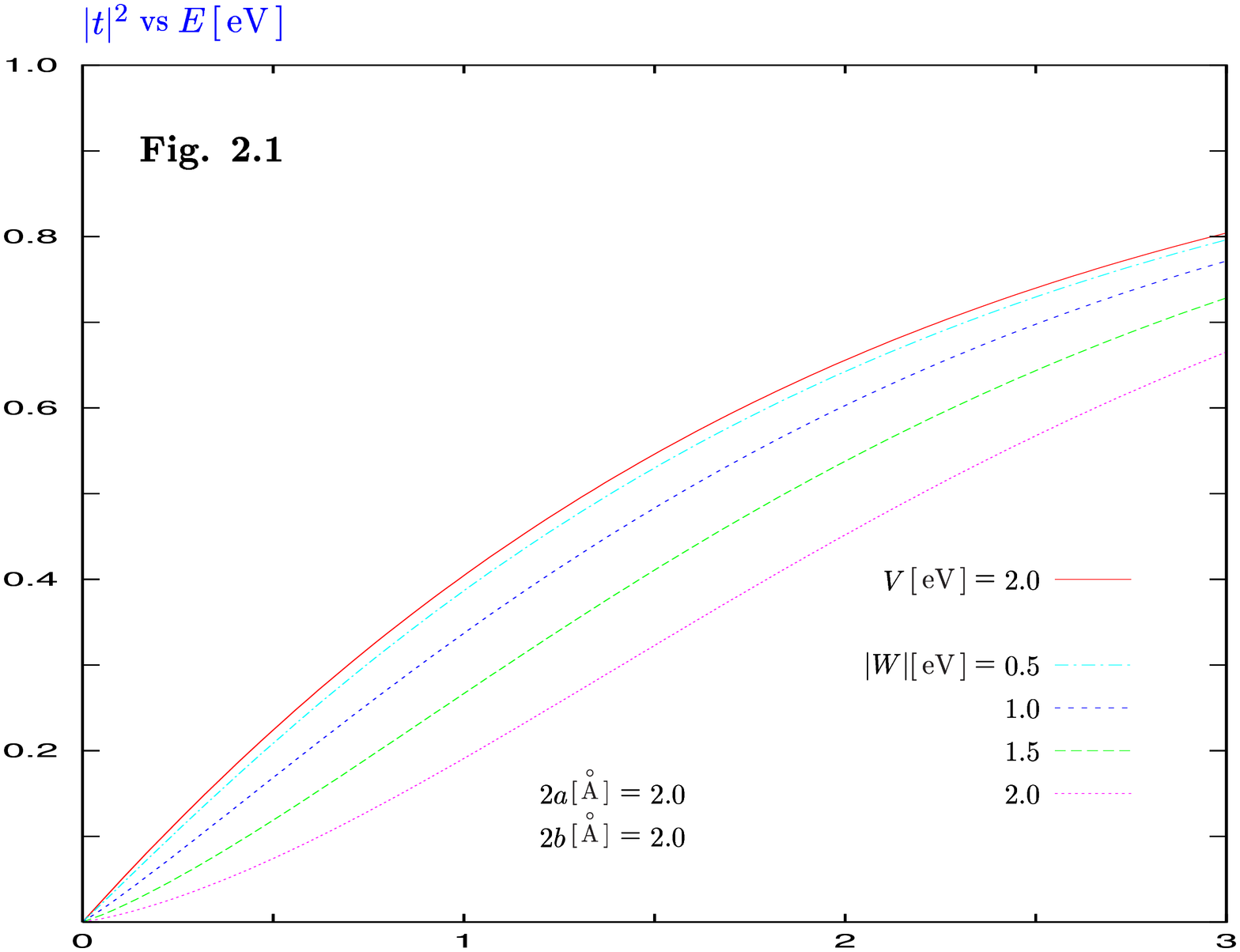}
\hspace*{-3.8cm}
\includegraphics[width=10.5cm, height=19.5cm, angle=90]{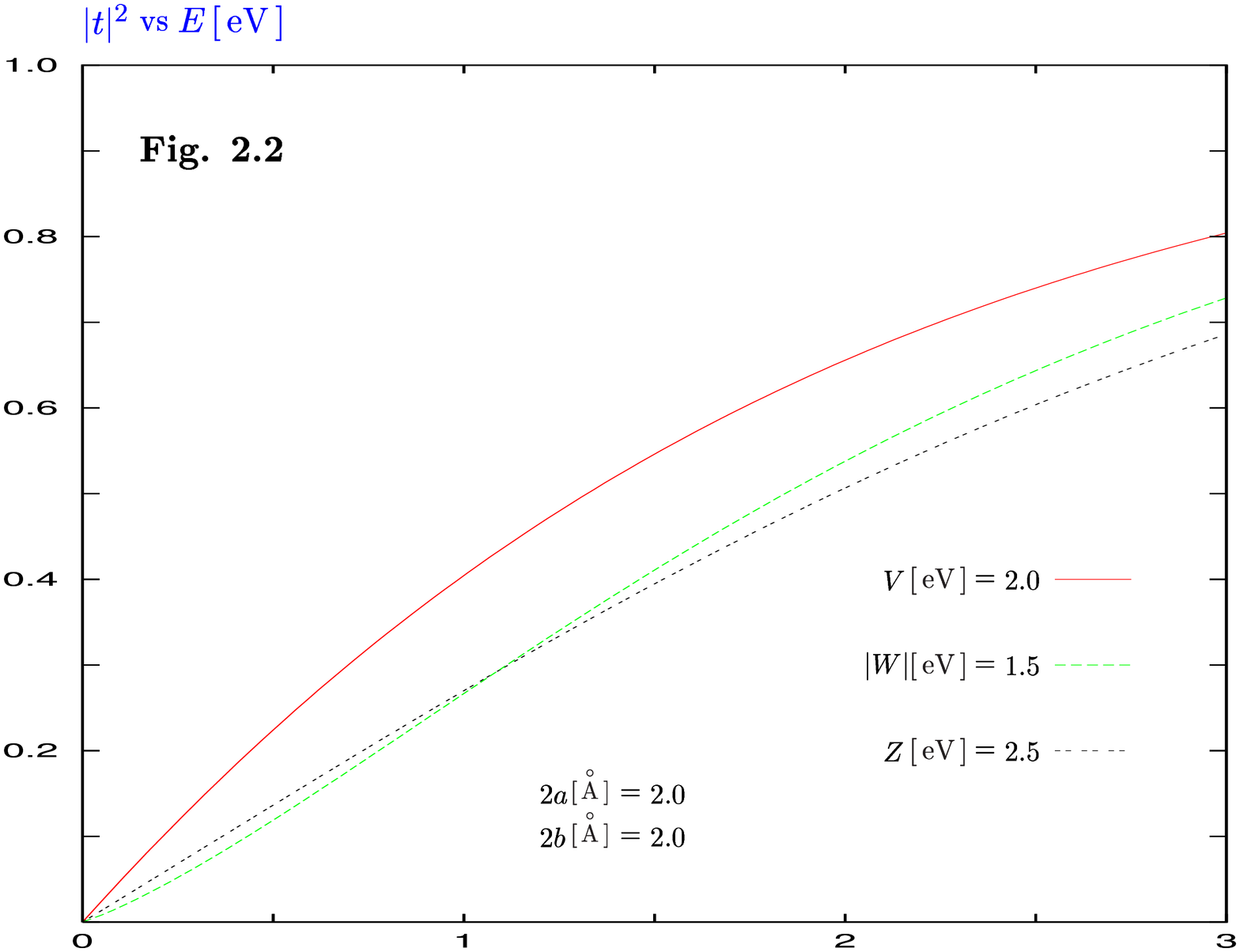}
\caption{Electron transmission probability, $|t|^{\2}$, as a function of 
$E$[eV] for quaternionic time reversal invariant potentials~\cite{DATA}.
The full line indicates the complex quantum 
mechanics result for the potential barrier of width 
$a$[{\tiny{$\overset{\circ}{A}$}}]$=1.0$ 
and height $V$[eV]$=2.0$. 
The dashed lines (drawn for a fixed width 
$b$[{\tiny{$\overset{\circ}{A}$}}]$=1.0$  and different values of the 
height $|W|$ of the potential $j \, W$)
show the quaternionic perturbation effects and the transmission 
probability for the complex
(comparative) barrier $Z=\sqrt{V^{\2}+|W|^{\2}}$.}
\end{figure}

\newpage

\begin{figure}
\hspace*{-3.8cm}
\includegraphics[width=10.5cm, height=19.5cm, angle=90]{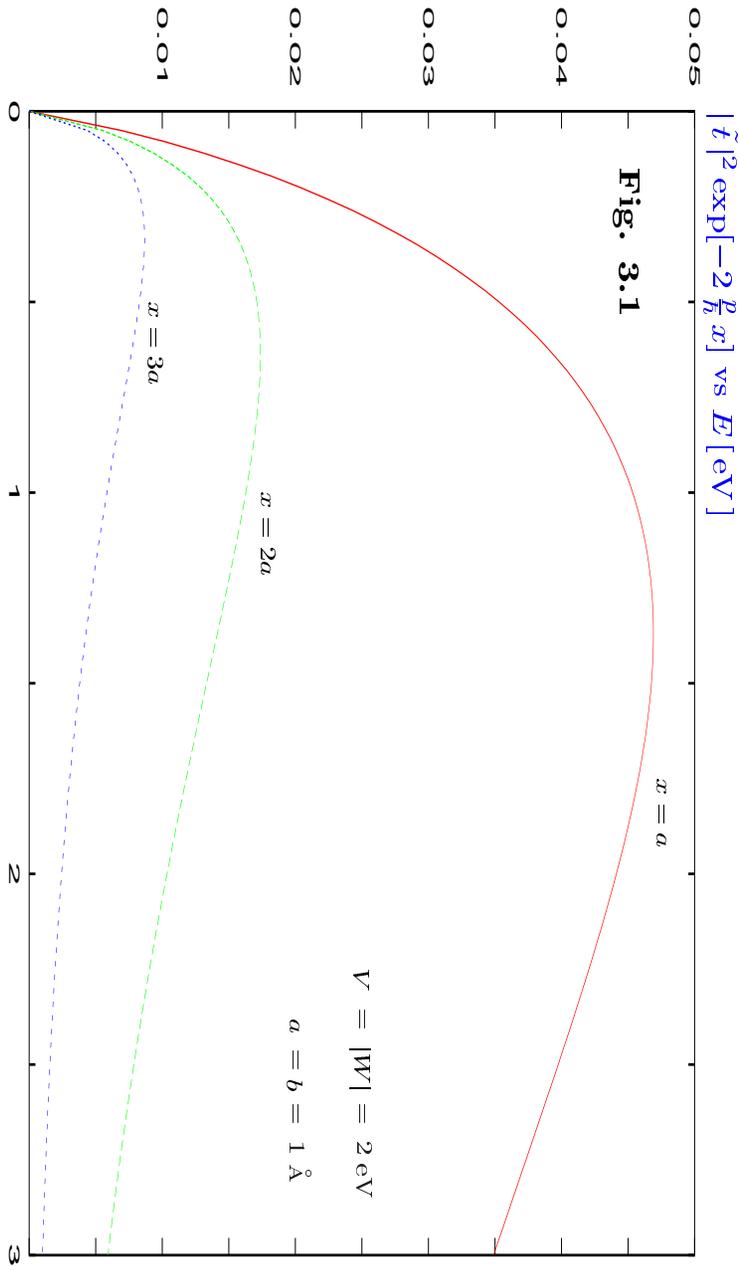}
\hspace*{-3.8cm}
\includegraphics[width=10.5cm, height=19.5cm, angle=90]{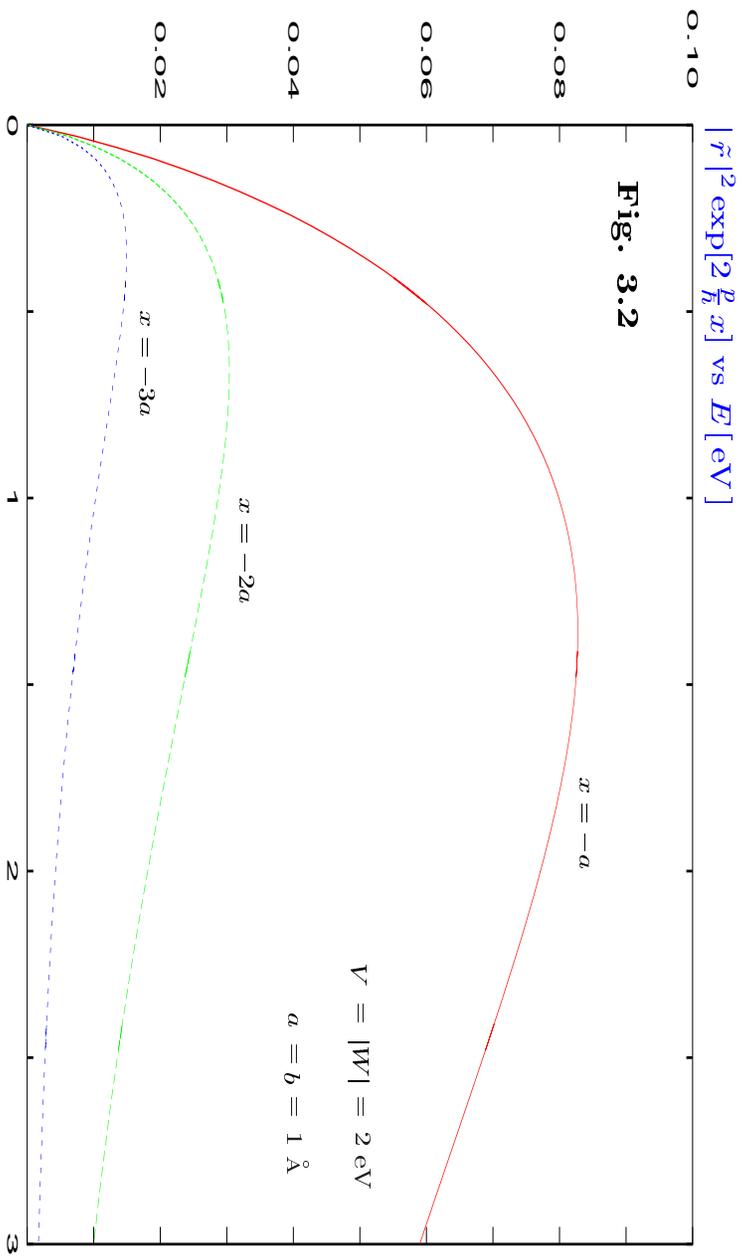}
\caption{Additional probability of electron transmission, 
$|\tilde{t}|^{\2} \exp[-2px/\hbar]$, and reflection,  
$|\tilde{r}|^{\2} \exp[2px/\hbar]$,  as a 
function of $E$[eV] for the quaternionic time reversal invariant potential
of width $a=b=1.0 \, \mbox{\tiny{$\overset{\circ}{A}$}}$ and height 
$V=|W|=2.0 \, \mbox{eV}$~\cite{DATA}. 
The curves show the additional
probability of transmission and reflection  for different values of $x$.}
\end{figure}

\newpage

\begin{figure}
\hspace*{-3.8cm}
\includegraphics[width=10.5cm, height=19.5cm, angle=90]{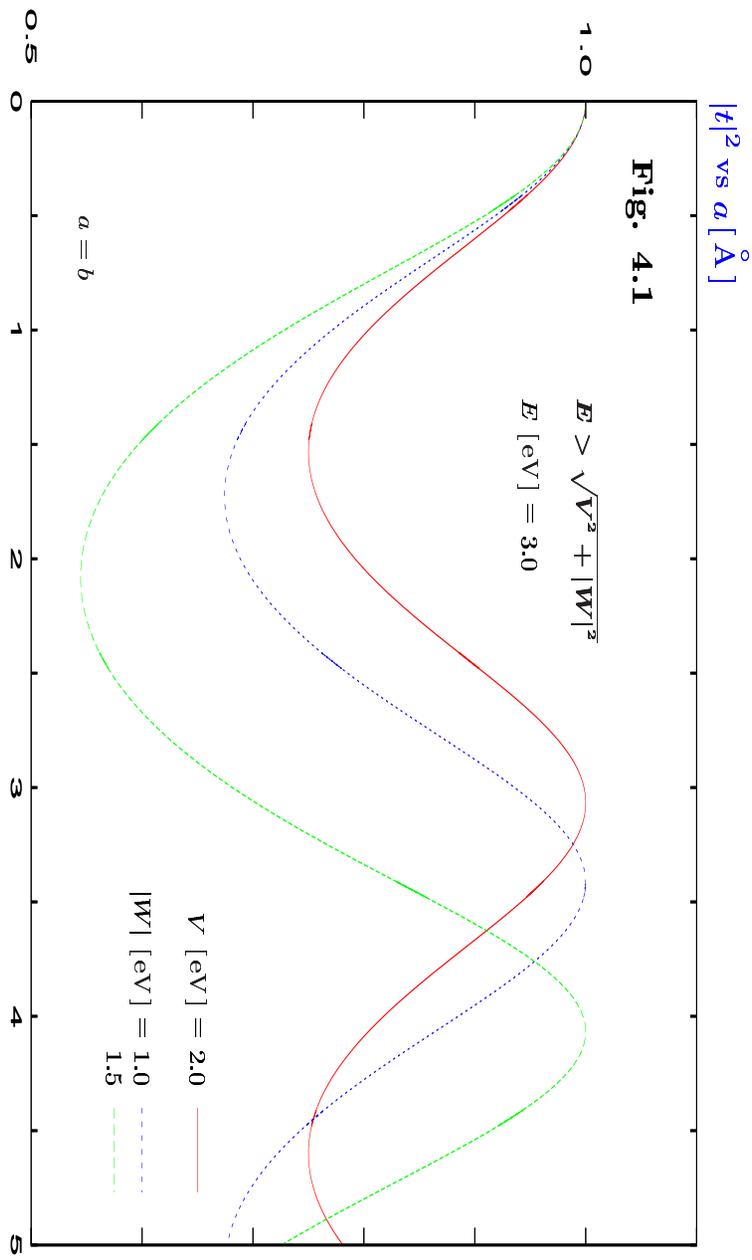}
\hspace*{-3.8cm}
\includegraphics[width=10.5cm, height=19.5cm, angle=90]{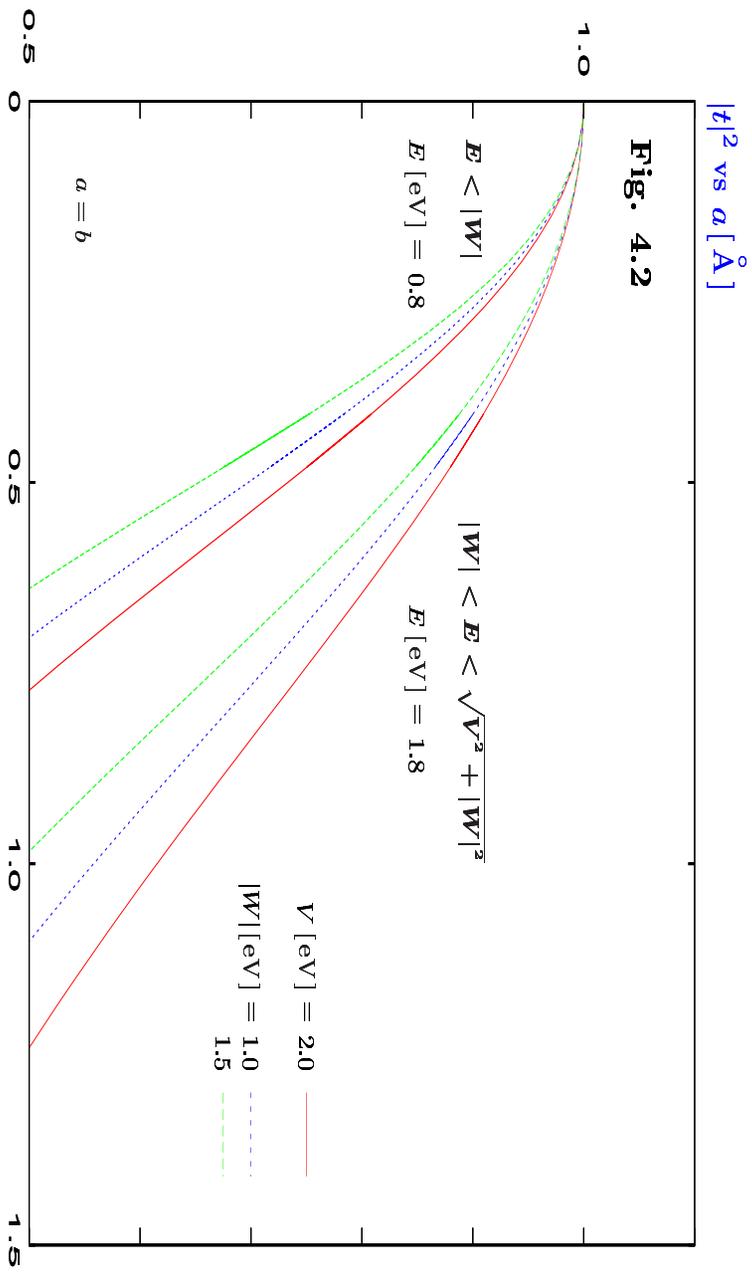}
\caption{Electron transmission probability, $|t|^{\2}$, as a function of 
$a$[{\tiny{$\overset{\circ}{A}$}}] for quaternionic time reversal invariant 
potentials~\cite{DATA}. The curves (drawn for different values of $E$) show 
the transmission probability  for the complex quantum mechanics  potential 
barrier of height  $V$[eV]$=2.0$ and for potentials of the same complex height
and quaternionic height $|W|$[eV]$=1.0$ and $1.5$.}
\end{figure}

\newpage

\begin{figure}
\hspace*{-3.8cm}
\includegraphics[width=10.5cm, height=19.5cm, angle=90]{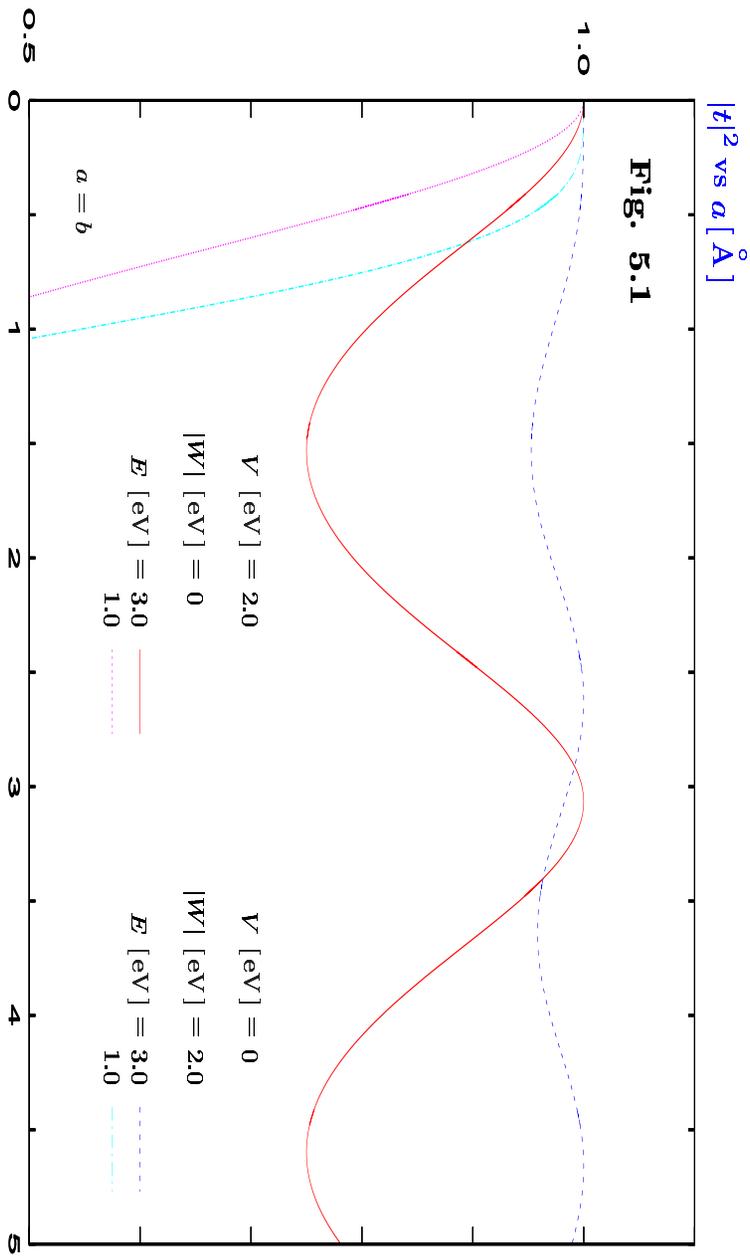}
\hspace*{-3.8cm}
\includegraphics[width=10.5cm, height=19.5cm, angle=90]{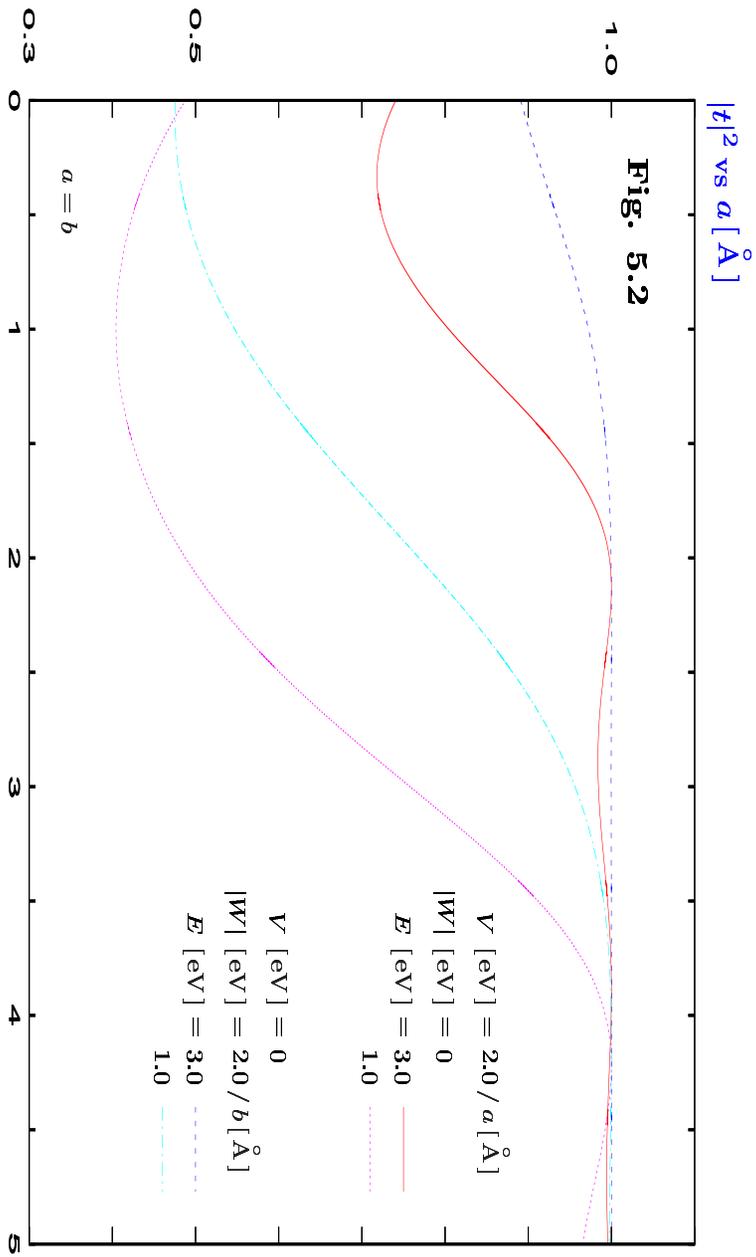}
\caption{Electron transmission probability, $|t|^{\2}$, as a function of 
$a$[{\tiny{$\overset{\circ}{A}$}}] for complex and {\em pure} quaternionic
potentials with the same width and height~\cite{DATA}. The curves show the 
transmission probability for different values of $E$.}
\end{figure}

\newpage

\begin{figure}
\hspace*{-3.8cm}
\includegraphics[width=10.5cm, height=19.5cm, angle=90]{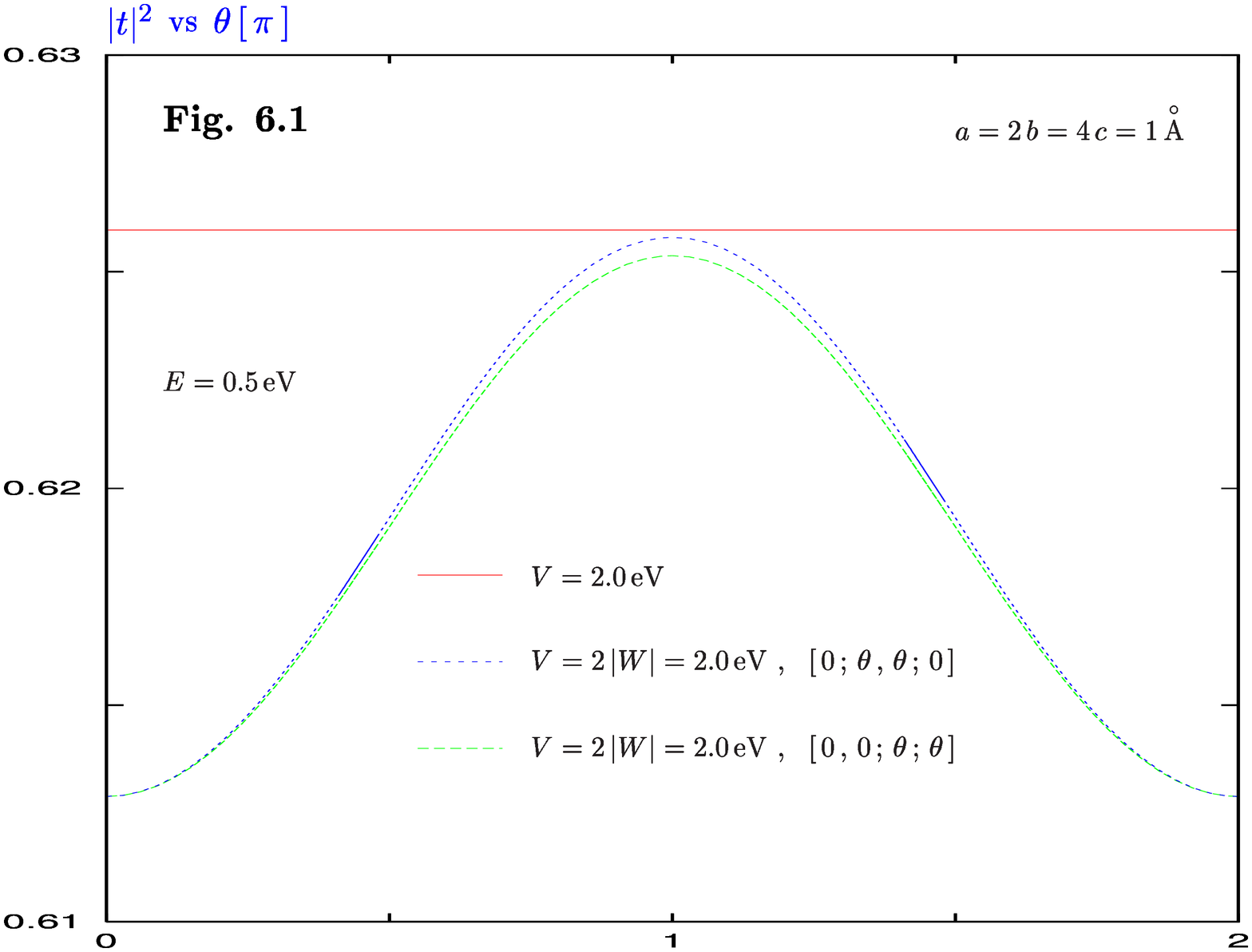}
\hspace*{-3.8cm}
\includegraphics[width=10.5cm, height=19.5cm, angle=90]{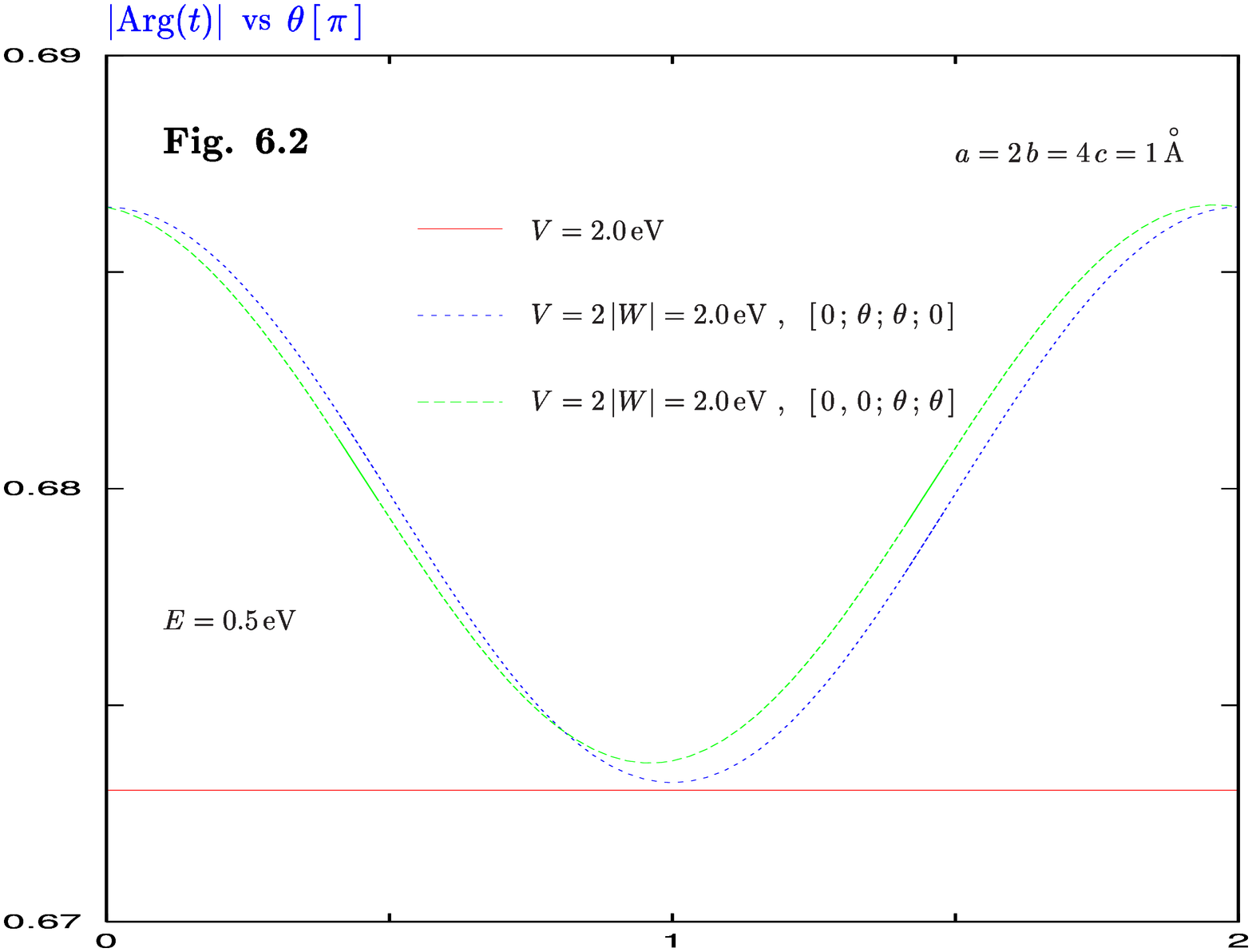}
\caption{Electron transmission probability, $|t|^{2}$, and absolute value of the
transmission coefficient argument, $|Arg(t)|$, as function of the
time violating phase $\theta[\pi]$ for potentials of height
$V=2|W|=2.0 \, \mbox{eV}$ and width 
$a=2b=4c=1.0\,${\tiny{$\overset{\circ}{A}$}}~\cite{DATA}.
The curves show that only 
asymmetric (time violating) quaternionic potentials could distinguish 
between left and right transmission. The value of the energy is fixed to
$E=0.5 \, \mbox{eV}$.}
\end{figure}

\newpage

\begin{figure}
\hspace*{-3.8cm}
\includegraphics[width=10.5cm, height=19.5cm, angle=90]{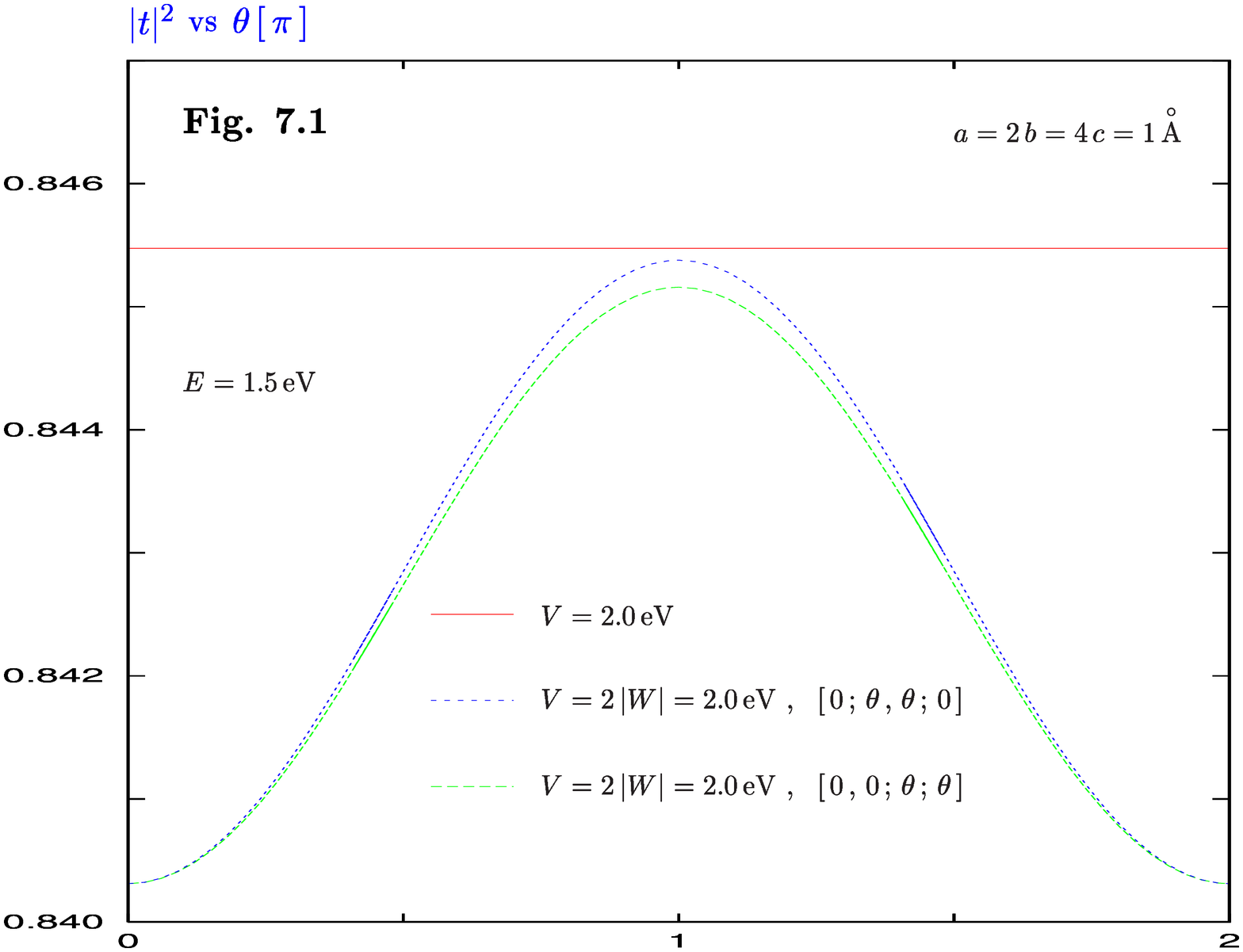}
\hspace*{-3.8cm}
\includegraphics[width=10.5cm, height=19.5cm, angle=90]{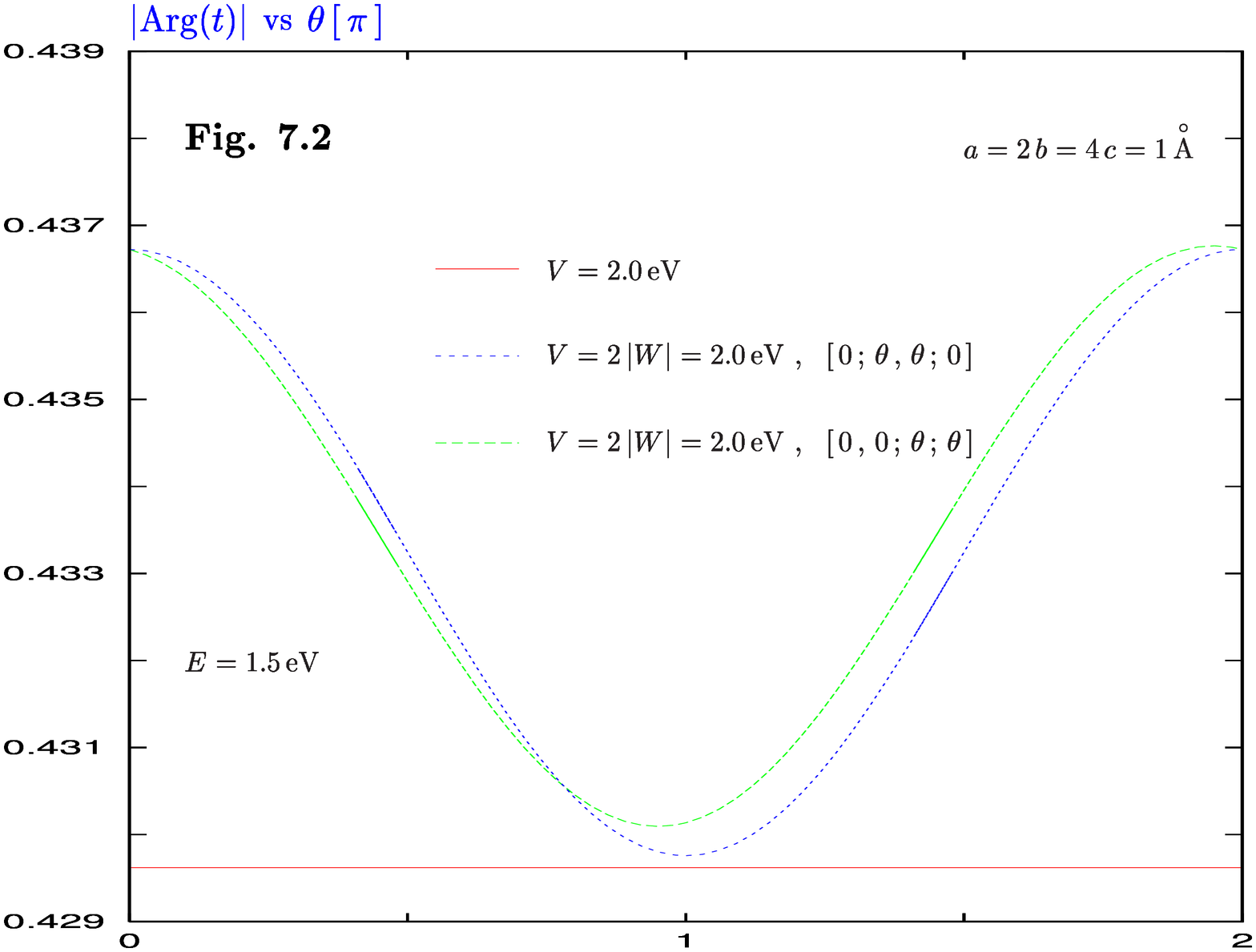}
\caption{Electron transmission 
probability, $|t|^{2}$, and absolute value of the
transmission coefficient argument, $|Arg(t)|$, as function of the
time violating phase $\theta[\pi]$ for potentials of height
$V=2|W|=2.0 \, \mbox{eV}$ and width 
$a=2b=4c=1.0\,${\tiny{$\overset{\circ}{A}$}}~\cite{DATA}.
The curves show that only 
asymmetric (time violating) quaternionic potentials could distinguish 
between left and right transmission. The value of the energy is fixed to
$E=1.5 \, \mbox{eV}$.}
\end{figure}

\newpage

\begin{figure}
\hspace*{-3.8cm}
\includegraphics[width=10.5cm, height=19.5cm, angle=90]{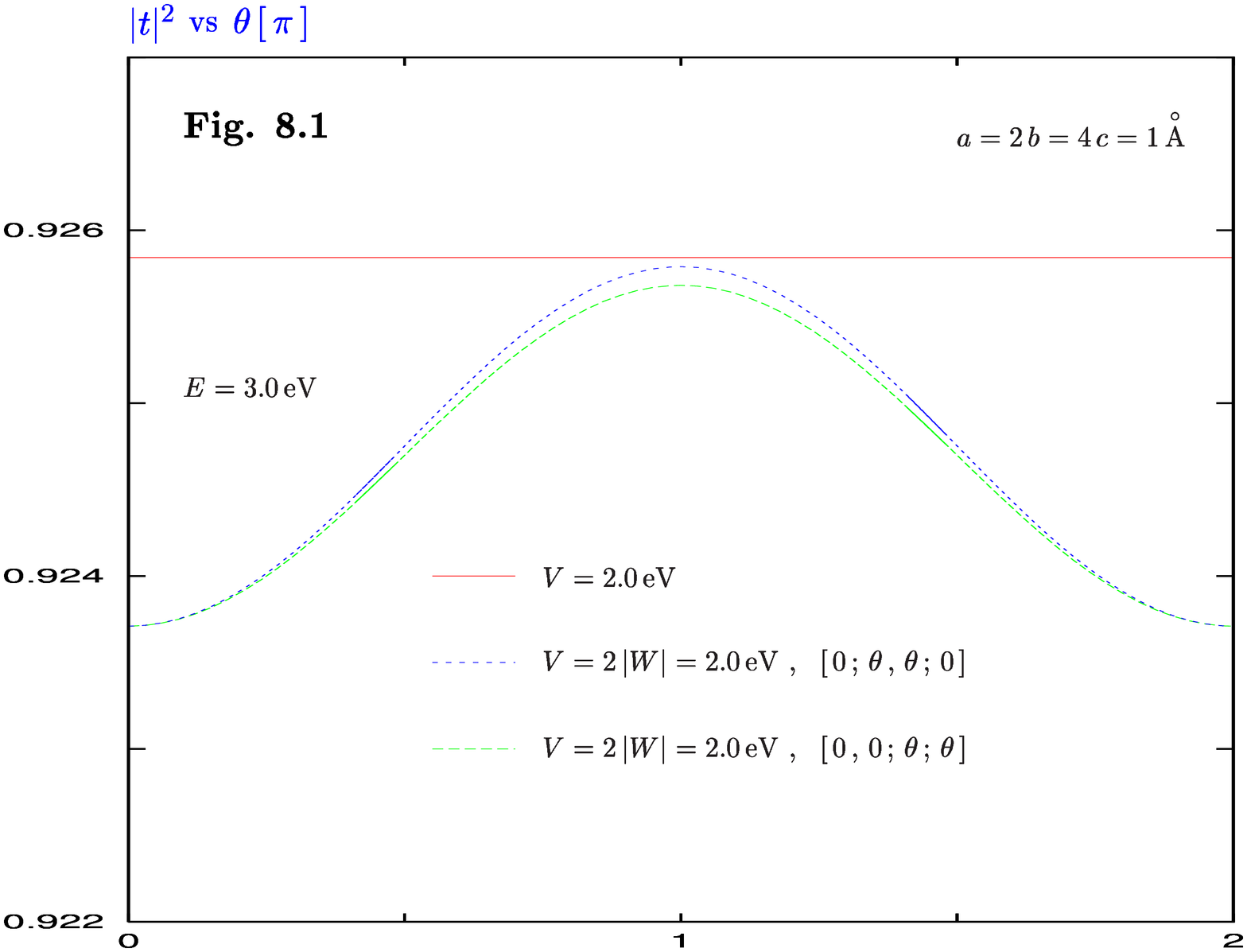}
\hspace*{-3.8cm}
\includegraphics[width=10.5cm, height=19.5cm, angle=90]{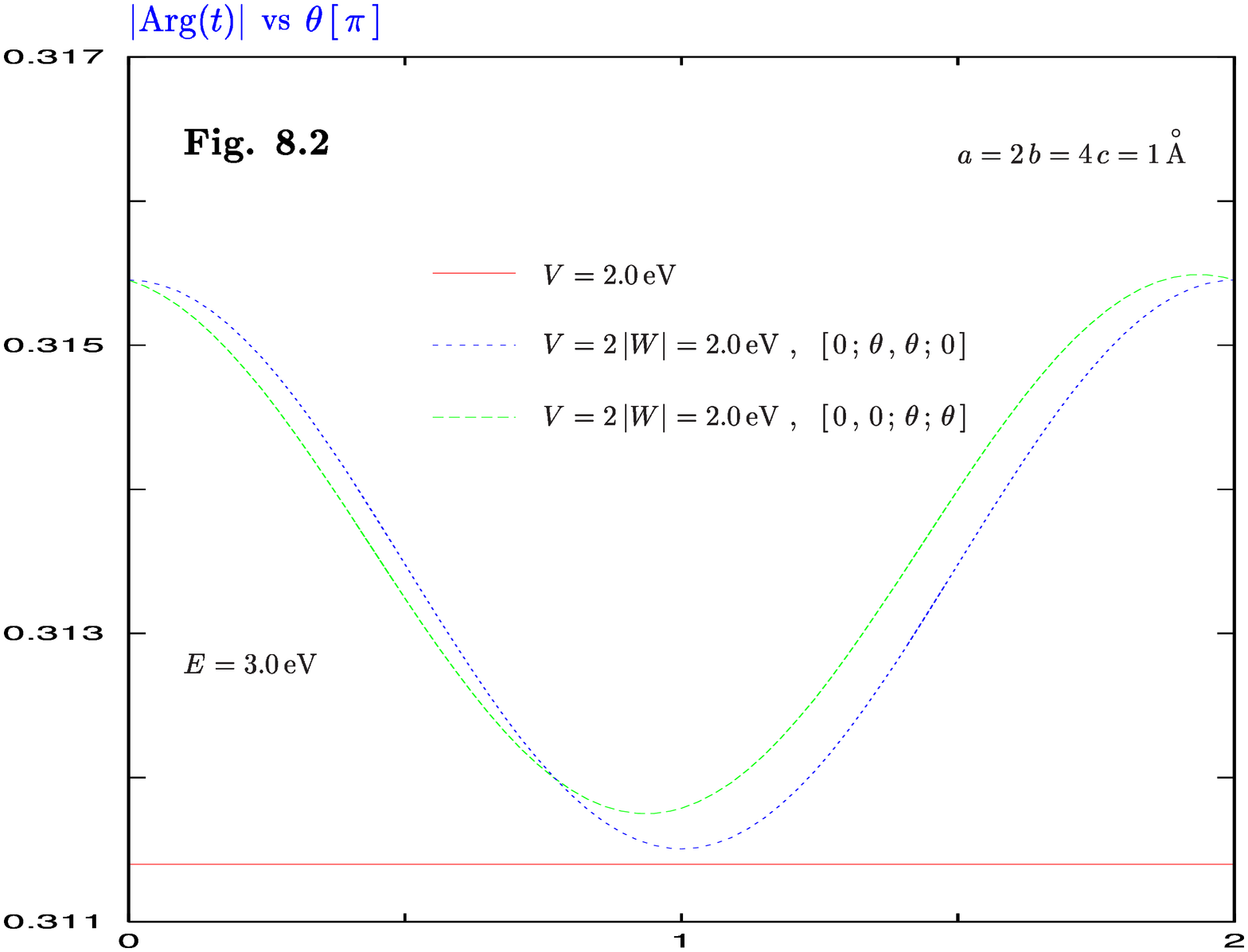}
\caption{Electron 
transmission probability, $|t|^{2}$, and absolute value of the
transmission coefficient argument, $|Arg(t)|$, as function of the
time violating phase $\theta[\pi]$ for potentials of height
$V=2|W|=2.0 \, \mbox{eV}$ and width 
$a=2b=4c=1.0\,${\tiny{$\overset{\circ}{A}$}}~\cite{DATA}.
The curves show that only 
asymmetric (time violating) quaternionic potentials could distinguish 
between left and right transmission. The value of the energy is fixed to
$E=3.0 \, \mbox{eV}$.}
\end{figure}


\newpage

\begin{thebibliography}{99}

\bibitem{KAN60}
T. Kaneno,
``On a possible generalization of quantum mechanics'',
Prog. Theor. Phys. {\bf 23}, 17--31 (1960).



\bibitem{FIN62}
D. Finkelstein, J. M. Jauch, S. Schiminovich and D. Speiser,   
``Foundations of quaternion quantum mechanics'',  
J. Math. Phys. {\bf 3}, 207--220 (1962).


\bibitem{FIN63}
D. Finkelstein and J. M. Jauch and D. Speiser,   
``Principle of general q covariance'',   
J. Math. Phys. {\bf 4}, 788--796 (1963).


\bibitem{EMC63}
J. Emch,   
``M\'ecanique quantique quaternionienne et relativit\'e restreinte'',  
Hel. Phys. Acta {\bf 36}, 739--788 (1963).



\bibitem{HOR84}
L. P. Horwitz and L. C. Biedenharn,
``Quaternion quantum mechanics: second quantization and gauge fields'', 
Ann. Phys. {\bf 157}, 432--488 (1984). 


\bibitem{ADL85}
S. L. Adler,
``Quaternionic quantum field theory'', 
Phys. Rev. Lett. {\bf 55}, 783--786 (1985). 

\bibitem{ADL86}
S. L. Adler,
``Super-weak CP non conservation arising an underlying quaternionic quantum 
dynamics'', 
Phys. Rev. Lett. {\bf 57}, 167--169 (1986). 


\bibitem{ADL86b}
S. L. Adler,
``Time-dependent perturbation theory for quaternionic quantum mechanics, 
with application to CP non conservation in K-meson decays'', 
Phys. Rev. D {\bf 55}, 1871--1877 (1985). 

\bibitem{ADL88}
S. L. Adler,
``Scattering and decay theory for quaternionic quantum mechanics and
the structure of induced T non-conservation'', 
Phys. Rev. D {\bf 37}, 3654--3662 (1988). 

\bibitem{ADL96}
S. L. Adler, 
``Generalized quantum dynamics as pre-quantum mechanics'',
Nucl. Phys. {\bf B473}, 199--244 (1996).  


\bibitem{ROT89}
P. Rotelli,
``The Dirac equation on the quaternionic field'',
Mod. Phys. Lett. A {\bf 4}, 993--940 (1989).


\bibitem{DEL92}
S. De Leo and P. Rotelli,   
``Quaternion scalar field'',            
Phys. Rev. D {\bf 45},  575--579 (1992). 




\bibitem{DIX}
G. M. Dixon, 
{\it Division algebras: octonions, quaternions, complex numbers and
               the algebraic design of physics}
(Boston: Kluwer Academic Publushers, 1994).

\bibitem{GUR}
F. G\"ursey and C. H. Tze, 
{\it On the role of division, Jordan and related algebras in 
               particle physics},              
(Singapore: World Scientific, 1996).

\bibitem{ADL}
S. L. Adler,
{\it Quaternionic quantum mechanics and quantum fields},
(New York: Oxford University Press, 1995).



\bibitem{PER79}
A. Peres, 
``Proposed test for complex versus quaternion quantum theory'',  
Phys. Rev. Lett. {\bf 42}, 683--686 (1979).

\bibitem{KAI84}
H. Kaiser, E. A. George and S. A. Werner, 
``Neutron interferometric search for quaternions in quantum mechanics'',  
Phys. Rev. A {\bf 29}, 2276--2279 (1984).


\bibitem{KLE88}
A. G. Klein, 
``Schr\"odinger inviolate: neutron optical searches for violations of 
quantum mechanics'',  
Physica B {\bf 151}, 44--49 (1988).


\bibitem{DAV89}
A. J. Davies and B. H. McKellar,           
``Non-relativistic quaternionic quantum mechanics'',  
Phys. Rev. A {\bf 40},  4209--4214 (1989). 

\bibitem{DAV90}
A. J. Davies,            
``Quaternionic Dirac equation'',   
Phys. Rev. D {\bf 41}, 2628--2630 (1990).

\bibitem{DAV92}
A. J. Davies and B. H. McKellar,           
``Observability of quaternionic quantum mechanics'',  
Phys. Rev. A {\bf 46}, 3671--3675 (1992). 

\bibitem{RCEE}
S. De Leo and G. Scolarici, 
``Right eigenvalue equation in quaternionic quantum mechanics'',  
J. Phys. A {\bf 33}, 2971--2995 (2000).


\bibitem{QDO}
S. De Leo and G. Ducati, 
``Quaternionic differential operators'',  
J. Phys. Math.  {\bf 42}, 2236--2265 (2001).


\bibitem{PER96}
A. Peres, 
``Quaternionic  quantum interferometry'', {\sf quant-ph/9605024}.


\bibitem{DATA}
Corresponding computer-readable data files may be found at 
{\sf http://www.ime.unicamp.br/$\tilde{~}$deleo/vqm}.


\end{thebibliography}
\end{document}